\documentclass[aps,superscriptaddress,twocolumn,twoside,floatfix,pra,a4paper]{revtex4-2} %nofootinbib
\usepackage{times}
\usepackage{epsfig}
\usepackage{amsfonts}
\usepackage{amsmath}
\usepackage{amssymb,amsthm}
\usepackage{color}
\usepackage{multirow}
\usepackage{braket}
\usepackage{bbm}
\usepackage{latexsym}
\usepackage{amsfonts}
\usepackage{mathrsfs}
\usepackage{natbib}
\usepackage{verbatim}
\usepackage{gensymb}
\usepackage{caption}
\usepackage{subcaption}
\usepackage{subcaption}
\usepackage{graphicx}

\usepackage[colorlinks=true,linkcolor=blue,citecolor=magenta,urlcolor=blue]{hyperref}
\allowdisplaybreaks

\begin{document}

%\title{Mass-independent nonclassicality of a mechanical oscillator}
\title{Mass-Independent Scheme to Test the Quantumness of a Massive Object}
%\title{Mass-independent quantumness of a mechanical oscillator}
%\title{Mass-independent nonclassicality of an macro-oscillators evidenced through  quantum violation of macrorealism}	

\author{Debarshi Das}
\email{debarshi.das@ucl.ac.uk}
\affiliation{Department of Physics and Astronomy, University College London, Gower Street, London WC1E 6BT, England,  United Kingdom}
\author{Dipankar Home}
%\email{dasdebarshi90@gmail.com}
\affiliation{Center for Astroparticle Physics and Space Science (CAPSS), Bose Institute, Kolkata 700 091, India}
\author{Hendrik Ulbricht}
%\email{dasdebarshi90@gmail.com}
\affiliation{School of Physics and Astronomy, University of Southampton, Southampton SO17 1BJ, England, United Kingdom}
\author{Sougato Bose}
%\email{dasdebarshi90@gmail.com}
\affiliation{Department of Physics and Astronomy, University College London, Gower Street, London WC1E 6BT, England, United Kingdom}

	\begin{abstract}
	The search for empirical schemes to evidence the nonclassicality of large masses is a central quest of current research.  However, practical schemes to witness the irreducible quantumness of an arbitrarily large mass are still lacking.  To this end, we incorporate crucial modifications to the standard tools for probing the quantum violation of the pivotal classical notion of macrorealism (MR): while usual tests use the same measurement arrangement at successive times, here we use  two different measurement arrangements. This yields a striking result:  a {\em mass-independent} violation of MR is possible for harmonic oscillator systems.  In fact, our adaptation enables probing quantum violations for literally any mass, momentum, and frequency.  Moreover, coarse-grained position measurements at an accuracy much worse than the standard quantum limit, as well as knowing the relevant parameters only to this precision, without requiring  them to be tuned, suffice for our proposal. These should drastically simplify the experimental effort in testing the nonclassicality of massive objects ranging from atomic ions to macroscopic mirrors in LIGO.
\end{abstract}
\maketitle

{\it Introduction and motivation:--} A cutting-edge research enterprise in contemporary physics is to explore realizable schemes for checking the validity of the quantum mechanics (QM) in the macroscopic regime, together with demonstrating its incompatibility with the world view based on the classical notion of macrorealism (MR) \cite{leggett02}. The goal is to expand as much possible the domain of such testability. This  also has potentiality in providing useful empirical constraints on suggested modifications of quantum dynamical evolution in the macroscopic limit (such as the  models of spontaneous wave function collapse \cite{swfc1,swfc2,BassiRMP}).   Nonclassical massive matter states are also a resource for nonclassical gravity  \cite{Bose2017, Marletto2017,Marshman2020,Bose2022,Rovelli2019,Christodoulou2022}.  Testing nonclassicality via MR can be, in principle, much easier than creating highly nonclassical states. However, in practice it imposes very high demands on the initial control of parameters,  as well as precise measurements, with its scaling becoming prohibitively difficult for large masses \cite{LHO1,LHO2}.  Here we show that appropriately modifying the schemes for testing MR in the context of massive objects  provides a threefold advantage: (i) We can obtain a  \textit{mass-independent} violation of MR, so that the applicability domain becomes essentially unlimited. (ii) It does not require any tuning of other parameters, e.g. frequency, momentum either. (iii) It becomes highly robust to measurement imprecision (no need to surpass the standard quantum limit, for example), opening up the scope for practical realizations. 

The key tools for probing MR are  ``temporal correlators" from which one constructs the Leggett-Garg inequality (LGI) \cite{lgi1} and the no-signalling-in-time (NSIT)  conditions \cite{nsit}. Such relations are derived from a conjunction of the following assumptions, as formulated by Leggett and Garg \cite{lgi1,lgi2} for characterizing the notion of MR: (i) At any instant, even if unobserved, a system is definitely in one of its possible states with all its observable properties having definite values (realism \textit{per se}). (ii) It is possible to determine which of the states the system is in by ensuring that the measurement-induced disturbance is arbitrarily small, and thus not affecting the subsequent time evolution of the measured state of the system (noninvasive measurability). (iii) The outcome of a measurement is not affected by what will be measured subsequently (induction). Since both the LGI and the NSIT relations are consequences of MR, an experimental refutation of either of them, in accordance with the quantum mechanical predictions, would constitute a decisive evidence of macroscopic nonclassicality, together with certifying the validity of the QM principle of superposition of states. While the logical connection between the LGI and NSIT has been analyzed in various ways \cite{explgi2,halliwelllginsit,KneePRA}, for the purpose of the present work it suffices to regard  violation of  either of them as a sufficient condition for evidencing nonclassicality or quantumness. From the operational point of view, using the NSIT relations is, in general, more advantageous compared to LGI because of the lower number of required outcome probabilities. Furthermore, since the NSIT condition is violated by the presence of any nonvanishing quantum interference term \cite{nsit}, it is usually violated for a much wider parameter regime than the LGI. %An experimental refutation of MR embodying these assumptions, in accordance with the QM predictions, would constitute a decisive evidence of macroscopic nonclassicality, together with certifying the validity of QM principle of superposition of states at the macrolevel.

With growing interest in this fundamentally significant topic, particularly over the past two decades, a number of experimental studies seeking to test MR in the macroscopic regime  have been reported (for a useful review of the earlier experiments, see Ref. \cite{qlgi1}). %Among the experiments in the recent years, the following ones need to be specially noted: (a) Violation of LGI has been shown using neutrino flavour oscillations \cite{explgi1} where the relevant ``macroscopicity'' measure is the length scale of neutrino oscillation of about 735 km over which the LGI violation has been found. (b) In the experiment involving a superconducting flux qubit \cite{explgi2}, the relevant ``macroscopicity'' measure is taken to be the difference of the magnetic moments corresponding to the two superposing superconducting current states (each of 170 nA current); this difference is of the order of $10^4$ Bohr magneton, and the empirical violation of MR is shown using a suitable NSIT condition. (c) In a very recent interferometric experiment based on heralded single photons \cite{explgi3}, loophole-free violations of both LGI and its variant \cite{wlgi} have been demonstrated, where the ``macroscopicity'' measure is of the order of $10^4$, specifying the extent to which the spatial separation between the two superposing single photon states (corresponding to the two arms of the interferometer) is larger than the photonic wavelength of 810 nm used in this experiment. It should therefore be clear from the preceding discussion that 
For characterizing ``macroscopicity'', these studies have used different parameters ranging from the length scale of neutrino oscillation \cite{explgi1} to the difference of the magnetic moments corresponding to the two superposing superconducting-current states \cite{explgi2} and  the spatial separation between the two superposing single-photon states (corresponding to the two arms of an  interferometer) \cite{explgi3}. However, while ``mass'' seems to be a quite relevant parameter for charaterizing ``macroscopicity'',   no experiment testing MR has yet been performed based on systems having significantly large mass - a few earlier LGI-based experiments employing atomic systems have been confined to using, for example, a single cesium atom \cite{explgi4}  and spin-bearing phosphorus impurities in silicon \cite{explgi5}. On the other hand, the tests of quantumness \textit{per se} of macromolecules (without seeking to test MR) have so far reached only up to masses of about $10^4$ amu \cite{macro1,macro2}. Our present work is motivated toward filling this important gap in the relevant literature by formulating a suitable scheme that can enable scaling up the test of MR \textit{vis-\`{a}-vis} QM to arbitrarily large masses of harmonically oscillating objects. To this end, we introduce a hitherto unexplored suitable variation of the LGI and the NSIT relations such that a  mass-independent QM violation of MR is possible. In fact,  for literally  any choice of parameters--mass,  initial momentum, frequency--our procedure can certify macroscopic quantumness and show violation of MR. It is an added advantage that the scheme, by working for highly imprecise measurements, provides a great facilitation of practical realization (in comparison to existing literature on large masses \cite{LHO1,LHO2}).

%\section{The basic ideas of our scheme}

{\it The basic ideas of our scheme:--} Let us begin by noting that the various versions of macrorealist inequalities/conditions that have been applied in different contexts usually consider the same observable to be successively measured on a single particle evolving in time. In contrast here, for the example considered, the successive measurements are invoked in such a way that they pertain to different observables. Let us now explain how this is realized. For a one-dimensional system harmonically oscillating between $x = - \infty$ and $x = \infty$, dividing this domain of oscillation into two subdomains, ranging from $x = - \infty$ to $x = \beta_i$ and from $x = \beta_i$ to $x = \infty$, where $\beta_i$ is any real number, we consider coarse-grained spatial measurement at an instant $t=t_i$ that determines which one of these two regions the oscillating system is in at the given instant. A key element of our scheme is that the location $x = \beta_i$ of the boundary between the two regions for the type of measurement considered is chosen according to the instant $t=t_i$ at which the measurement is made.  For convenience, considering that the initial coherent-state Gaussian wave function at $t=0$ is peaked at $x=0$,  we choose the instant of the first measurement to be the initial instant, i.e., $t_1 = 0$ and we determine the boundary for this measurement to be located at $x=\beta_1 = 0$.

Next, for the subsequent measurement at the instant $t=t_2$, the appropriate choice of the location $x = \beta_2$ of the boundary between the two measurement regions is critical in order to achieve the desired mass independence of the quantum violation of MR. Toward attaining this goal, our analysis reveals that by suitably fixing $\beta_2$ for given values of $t_2$, it is possible to ensure the magnitudes of the quantum violations of both  the two-time LGI \cite{lg2} and the two-time NSIT relation \cite{nsit,wlgi} to be independent of mass, as well as of the other relevant experimental parameters such as the initial peak momentum and the angular frequency. Here it needs to be pointed out that if the observable quantity is taken to be such that at any instant, it takes a value $+1$ ($-1$) depending on whether the system is in one region or in the other, it is then evident that in this example, such an observable quantity changes according to the location of the boundary demarcating the two measurement regions. Hence, for the purpose of the following analysis, the two-time LGI and the two-time NSIT relation  invoked are, crucially, in terms of different observable quantities being measured at two different instants. Here note that the derivations of LGI and NSIT relations from MR do not depend upon the measured quantity  necessarily being the same at the different instants of successive measurements.

%A couple of further relevant points also require to be mentioned here. First, it is operationally advantageous to test two-time macrorealist relations compared to the three-time ones because of the lesser number of  joint probabilities measured. Among the two-time macrorealist relations, the NSIT condition involves minimum number of observable probabilities. Secondly,  

    An important point to note here is that the measurement envisaged in our example can be designed such that an outcome is inferred when the detector is not triggered (negative result measurement). This ensures that there is no classical interaction with the measured object during measurement, thereby satisfying the assumption of noninvasive measurability, if the measured object is classical \cite{lgi2}. Such a measurement can be implemented in our setup by using, say, a probe beam illuminating one of the two regions (either from $x = \beta_i$ to $x = \infty$, or from $x = \beta_i$ to $x = - \infty$).  If no scattered light is observed, an outcome is registered by inferring the presence of the oscillating object within the unilluminated region \cite{LHO1}. Other relevant specifics will be discussed with respect to the analysis of our scheme presented as follows.

\textit{The analysis:--} We begin by explicitly writing the modified forms of the two-time LGI  and the two-time NSIT relation involving sequential measurements of the observables denoted by $Q$ and $R$ at two different instants $t=t_1$ and $t=t_2$ respectively,  where $t_1 < t_2$. The NSIT condition implies that the probability of
obtaining a particular outcome for the measurement
of $R$ at $t=t_2$ should be independent of whether any prior measurement has been carried out. That is, it can be regarded as the statistical version of noninvasive measurability. In the present context, the two-time NSIT condition can be expressed as
\begin{align}
	&\text{N}_{\pm}  =	p(R\pm) - \left[  p(Q+, R \pm) +  p(Q-,R\pm) \right] =0,
	\label{nsit}
\end{align}
where $p(Q\pm, R \pm)$  is the joint probability of getting the outcomes $\pm 1$ at instant $t=t_1$ and $\pm 1$ at  instant $t=t_2$; $p(R \pm)$ is the probability of getting the outcome $\pm 1$ at  instant $t=t_2$, when no measurement is done at $t=t_1$. The magnitude of the quantum violation of the two-time NSIT condition will be denoted by the nonzero value of $|\text{N}_{\pm}|$.

In this scenario, the two-time LGI can be expressed as \cite{supp}
\begin{align}
	\text{L}_{s_1,s_2} = &1 + s_1 \langle Q \rangle + s_2 \langle R  \rangle + s_1 s_2 \langle QR \rangle \geq 0, \nonumber \\
	&\text{with} \, \, \,  s_1, s_2 \in \{+1, -1\},
\end{align}
where the correlation function is $\langle QR \rangle = p(Q+, R+) - p(Q+, R-) - p(Q-, R+) + p(Q-, R-)$; and the expectation values are $\langle Q \rangle = p(Q+) - p(Q-)$, $\langle R \rangle = p(R+) - p(R-)$. Here $\langle R \rangle$ is defined when no measurement at $t=t_1$ is performed. In this case, magnitude of quantum violation of the two-time LGI will be denoted by the positive value of $\max\limits_{s_1 = \pm 1, s_2 = \pm 1} (- \text{L}_{s_1,s_2} )$.

%One point to be stressed here is that though the LGI and the NSIT condition test different notions of MR, as argued recently \cite{halliwelllginsit}, we treat their violations on equal footing as both are sufficient conditions for witnessing non-classicality or quantumness.

Consider the following initial Gaussian wave function of the coherent state
peaked at $x=0$ at instant $t=0$:
\begin{align}
	\psi(x,t=0) &= \sqrt{\frac{1}{\sqrt{2 \pi} \sigma_0}}  \text{exp}\left(-\frac{x^2}{4 \sigma_0^2} + \frac{\mathbbm{i} p_0 x}{\hbar}\right)
\end{align}
with the initial momentum expectation value $p_0$, and the
width $\sigma_0 = \sqrt{\hbar/(2 m \omega)}$, where $\omega$ is the angular frequency of oscillation and $m$ is the mass. The time evolution of this state is evaluated in this example using linear harmonic oscillator potential.

We consider measurements of $Q$ and $R$ at the instants $t=t_1$ and $t=t_2$ respectively, where $Q$ and $R$ correspond to the earlier-mentioned coarse-grained measurements. To put it specifically, $Q$
is an observable quantity such that  it takes a
value $+1$ ($-1$) depending on whether the system is in the
region from $x=\beta_1$ to $x=\infty$ (from $x=-\infty$ to $x=\beta_1$). Similarly, $R$
is another observable quantity such that it takes a
value $+1$ ($-1$) if the particle is in the
region from $x=\beta_2$ to $x=\infty$ (from $x=-\infty$ to $x=\beta_2$). Such a coarse-grained position measurement at instant $t=t_i$ ($i = 1,2$) can be represented by the operator $\hat{O}_i = \int_{\beta_i}^{\infty} |x\rangle \langle x| dx - \int_{-\infty}^{\beta_i} |x\rangle \langle x| dx$, which has two eigenvalues: $+1$ and $-1$. 

As mentioned earlier, $\beta_1=0$.  With this choice of $\beta_1$, it can be shown that  the expressions for the joint probabilities are the following functions of $\beta_2$ and other relevant parameters:
\begin{align}
	&p(Q\pm,R-) = \frac{1}{4 \sqrt{\pi}} \int_{-\infty}^{\gamma} dx \, \text{exp}\left[ - x^2 \right] f(x, \omega t_2), \nonumber \\
	&p(Q\pm,R+) = \frac{1}{4 \sqrt{\pi}} \int_{\gamma}^{\infty} dx \, \text{exp}\left[ - x^2 \right] f(x, \omega t_2), \label{jointprm} \\
	&\text{with} \, \, f(x, \omega t_2) =  \left| 1 \pm \text{erf}\left[ \frac{ -\mathbbm{i} x}{\sin (\omega t_2) \sqrt{2 - 2 \mathbbm{i} \cot (\omega t_2) }} \right] \right|^2,  \\
	& \hspace{1cm} \text{and} \, \, \gamma = \beta_2 \sqrt{\frac{m \omega}{\hbar}} - \frac{p_0 \sin (\omega t_2)}{\sqrt{\hbar m \omega }},
	\label{gamma}
\end{align}
where the error function $\text{erf}(z) = (2/\sqrt{\pi}) \int_{0}^{z} dt \, \text{exp}(-t^2)$. Similarly, we have the following form of probabilities:
\begin{align}
	p(Q \pm) = \frac{1}{2}; \, \, \, \,	p(R \pm) =\frac{1}{2}\left[1\mp \text{erf}\left(\gamma\right) \right].
	\label{singleprm}
\end{align}
The details of these calculations are given in the Supplemental Material.

Next, we have to choose $\beta_2$ suitably for the measurement at $t=t_2$. Our heuristic arguments based on physical ground \cite{supp} suggest that the location of the peak at the instant $t=t_2$, given by $x^{(t_2)}_0 = p_0 \sin (\omega t_2)/(m \omega )$, together with the standard deviation $\Delta^{(t_2)} = \sqrt{\hbar /(2 m \omega)}$ of the probability density without any prior measurement at $t=t_1$, would play a critical role in fixing $\beta_2$ suitably for our purpose. Guided by this consideration, our in-depth investigation reveals that if the choice of $\beta_2$ is made at $x_0^{(t_2)} \pm c \Delta^{(t_2)}$ with $c$ being positive and of the order of $10^{-1}$ or $1$, the possibility indeed arises for obtaining quantum violations of both the two-time NSIT condition and the two-time LGI. Here a key point is that this choice of $\beta_2= x_0^{(t_2)} \pm c \Delta^{(t_2)}$ leads to $\gamma$ given by Eq.(\ref{gamma}) becoming independent of $m$, $\omega$ and $p_0$, i.e., $\gamma$ is then determined only by  the chosen value of $c$, whence $\gamma = \pm c/\sqrt{2}$. Consequently, the probability distributions (\ref{jointprm}) and (\ref{singleprm}) become functions of $(\omega t_2)$ only. This, therefore, enables the quantum violations of the two-time NSIT condition and the two-time LGI to become independent of mass. That this is indeed the case is confirmed comprehensively by evaluating  numerically the integrations appearing in (\ref{jointprm}). In Table \ref{tab1}, some illustrative results are presented by choosing, for example, $c= \sqrt{2}$. %These results indicate and is confirmed by detailed study that the two-time NSIT condition is violated for a wider range of choices of $t_2$ than that for the two-time LGI. 
To sum up, the upshot of this entire study is that it is always, in principle, possible to choose $t_2$ suitably depending on the time period of the oscillating particle to obtain quantum violation of the two-time NSIT condition or the two-time LGI for any $m$, $p_0$ and $\omega$.

{\centering
	\begin{table}
		\begin{tabular}{ |c|c|c|c|c|c| } 
			\hline
			&  &  &  & Magnitude of & Magnitude of  \\
			&  &  &  & quantum violation & quantum violation  \\
			&  &  &  & of two-time NSIT: & of two-time LGI: \\
			$m$ & $p_0$ & $\omega$ & $t_2$ & $|\text{N}_{\pm}|$  & $\max\limits_{s_1 = \pm 1, s_2 = \pm 1} (- \text{L}_{s_1,s_2})$ \\
			\hline
			\hline
			Any & Any  & Any & $\frac{T}{14}$ & 0.12 & 0.08  \\
			\hline
			Any & Any  & Any & $\frac{T}{8}$ & 0.15 & 0.04 \\ 
			\hline
			Any & Any  & Any & $\frac{T}{4}$ & 0.17 & No violation \\ 
			\hline
			Any & Any  & Any & $\frac{T}{3}$ & 0.16 & No violation \\ 
			\hline
			Any & Any  & Any & $\frac{3T}{8}$ & 0.15 & 0.04 \\ 
			\hline
			Any & Any  & Any & $\frac{2T}{5}$ & 0.14 & 0.07 \\ 
			\hline
			Any & Any  & Any & $\frac{3T}{4}$ & 0.17 & No violation \\ 
			\hline
		\end{tabular}
		\caption{Quantum violations of the two-time NSIT condition  and the two-time LGI when the boundary between the two regions in the case of the second measurement is chosen to be located at $x = \beta_2 = p_0 \sin (\omega t_2) /(m \omega ) + \sqrt{\hbar /(m \omega)}$. Here $T = 2 \pi / \omega$ denotes the time period.} \label{tab1}
	\end{table}
}

{\it Practical challenges with large mass and measurement precision:--} Ideally, in our scheme, one of the two regions (either from $x=-\infty$ to $x = \beta_i$ or  from $x= \beta_i$ to $x=\infty$) should be illuminated at an instant $t=t_i$. In practice, however, it is almost impossible to keep the boundary between the two regions (illuminated and unilluminated) fixed in all experimental runs. Rather, we can expect that the aforementioned boundary at $t=t_i$  will be at $x=\beta_i + \epsilon_i$ (with $\epsilon_i$ being a small positive/negative number depending on the accuracy of the experimental setup), where $\epsilon_i$ will be different in different runs.   In effect, the observed violation of the NSIT condition will be the statistical average over different values of $\text{N}_+$, $\text{N}_-$ corresponding to different values of $\epsilon_1$ and $\epsilon_2$. Similar will be the case for the  LGI. It can be shown that the permissible ranges of $\epsilon_1$ and $\epsilon_2$ to get significant violation of the NSIT or the LGI are proportional to $1/\sqrt{m \omega}$ \cite{supp}. Hence, the required precision in fixing the boundary between the two measurement regions at any instant is increased with increasing mass. Nonetheless, the effect of increasing mass in this context can be offset by decreasing the angular frequency $\omega$. Here, it is relevant to note that the lowest angular frequency of a harmonic well achieved to date is $\omega \sim 100$ kHz in case of an optical trap \cite{ot}, $\omega \sim 100$ Hz in an ion trap \cite{at1,at2} and 1-10 Hz in a diamagnetic trap \cite{dt1,dt2,microhertz}. For small $\omega$, the violations as mentioned in Table \ref{tab1} are observed for large $t_2$. For any $\omega$, damping has to be controlled so that the decoherence rate due to all force noises can be given by $\gamma = S_{FF}(\omega) (\Delta x)^2/\hbar^2 \ll 1/t_2$, where $S_{FF}(\omega)$ is the noise power density of force and $\Delta x$ is the width of the wave function at $t_2$. Since both \textit{with and without} measurement at $t_1$, $\Delta x \sim \sigma_0 = \sqrt{\hbar/(2 m \omega)}$ (see Figs. 1 and 2 in \cite{supp}), the above condition reduces to $\sqrt{S_{FF}(\omega)} \ll \omega \sqrt{\hbar m} /\sqrt{\pi}$ for $t_2 \sim T$ (for example, with $m \sim 10$ kg and $\omega \sim 100$ Hz \cite{Chris}, $\sqrt{S_{FF}}\sim 10^{-15} \text{N}/\sqrt{\text{Hz}}$ is required). Interestingly, large masses indeed help here, for the obvious reason that force noise induces less random acceleration. The above consideration of decoherence is quite generic: all recoil noises, as well as all trapping noises, can be encompassed under the above limit on force noise.

It is well known that the balance between unwanted measurement backaction and the
precision of optical measurements imposes a standard quantum limit (SQL) on the position measurement of a harmonically trapped object \cite{sql1}. For the system considered by us, this limit is given by, $\delta x = \sqrt{\hbar/(2 m \omega)}$ -- the minimum uncertainty in position measurement \cite{sql1,sql2}. Recently, it has been shown that one can surpass SQL, but only with highly delicate technologies \cite{sql3,sql4}. In our proposed setup, probing the quantum violation of MR for any $m$, $\omega$ and $p_0$ using the NSIT condition is possible even when the accuracies of the coarse-grained position measurements are much worse than in the SQL \cite{supp}.  However, this is not the case for the LGI. Hence, the NSIT condition should be preferred for implementing our proposal.

{\it Possible experimental implementations:--}  We can envisage implementations with nano- and micro-objects (typically up to $\sim 10^{-14}$ kg) based on so-called levitated mechanics~\cite{gonzalez2021levitodynamics} in various low-noise traps such as optical dipole traps, ion traps, and magnetic and diamagnetic traps in vacuum and at low temperature, as well as using much larger masses (e.g., $\sim$ mg \cite{matsumoto1, matsumoto2} and $\sim 10$ kg in  the gravitational wave detectors \cite{Chris}). The mass independency  of this MR test can be judiciously made use of  in choosing the experimental setups optimized to reduce relevant decoherence effects, and noises, as well as for addressing at the same time the need for high spatial detection resolution of the centre-of-mass motion of the trapped particle. Specifically, one can observe the quantum violations of MR as described in Table \ref{tab1} for any given mass dependent only on $t_2$ even if the values of $\omega$ and $p_0$ are different in different experimental runs--i.e., $\omega$ and $p_0$ need not be tuned in each run. The only requirement is that the values of $\omega$ and $p_0$ in each run need to be known in order to fix $\beta_2$. In particular, for testing the NSIT condition,  it is sufficient to know these parameters to the precision of the order of SQL. On the contrary, for LGI, these parameters should be known with much more precision implying less difficulty with testing the NSIT condition.

The preparation of the initial state will be accomplished by cooling, for instance, by feedback~\cite{sub-kelvin-Giesler}, to a motional state of low occupation number,  for which the only requirement is that the rate of acquiring information  about the object must be much faster than the rate of its heating from environmental noise \cite{Doherty-Jacobs,Daley,Ulbricht}.  This technique has already been used for cooling to the ground state for $\omega \sim 100$ kHz traps \cite{Magrini-Aspelmeyer,Tebbenjohanns-Novotny} as well as for a large-mass and low-frequency ($\omega \sim 100$ Hz)  gravitational wave detector~\cite{Chris}. The conditions  are well within ultrahigh vacuum at $10^{-10}$~mbar and can be fulfilled in low-temperature environments even below 10~mK and with vibration isolation. %{\color{red} We consider the high efficiency and high spatial resolution of detection the most severe challenge for a realisation of this MR scheme, similar as for measurement-based state manipulation schemes~\cite{Kurt, Gerd}. 

While keeping the trapping a low-noise mechanism (e.g., an ion trap, magnetic, or diamagnetic trap, etc.), a promising possibility for the realization of measurements in this experiment is optical detection. If we are using the detection of scattered light from the particle to detect its motion, for an assumed near-unity efficiency of collection of the scattered light, and for $n$ photons,  we obtain a resolution of $\lambda/\sqrt{n}$ with $\lambda \sim \mu$m. This implies the collection of $n\sim  10^8$ photons for reaching a spatial resolution of the ground-state spread $\sigma_0$ of a $m\sim 10^{-14}$kg particle in a $\omega \sim 1$ Hz trap.  If the detected scattered power is 1 nW, this information is acquired in $\sim 10^{-2}$ s, implying that all other heating rates, such as undetected scattered photons, from interactions with blackbody photons and gas collisions have to be at $\Gamma \leq 10^{2}$ Hz.    

For increasing the critical detection efficiency in light-scattering techniques, one could use collection optics with a high numerical aperture--for example,  parabolic mirrors~\cite{Vovrosh-Ulbricht}. The high spatial resolution could be achieved by illuminating only either to the left or right of $x=\beta_i$ at instant $t=t_i$ with a sharp drop of profile  at the point $x=\beta_i$. If no scattered light is obtained after $0.01$ s with a nW laser illuminating the left half,  it immediately implies a $+1$ outcome, with the location $x=\beta_i$ being at angstrom resolution. Such spatial resolution has been achieved in the optical imaging of single molecules~\cite{Vahid} and in optomechanical experiments~\cite{Ulbricht,Tracy} by using optical interferometry. 

%For increasing the critical detection efficiency in light scattering techniques one could use collection optics with high numerical aperture, for example,  parabolic mirrors~\cite{Vovrosh-Ulbricht}. The high spatial resolution could be achieved by illuminating only either to the left or right of $x=\beta_i$ at instant $t=t_i$ with a sharp drop of profile  at the point $x=\beta_i$. If after $0.01$ s with a nW laser,  no scattered light is obtained,  it immediately implies that the operator having acted upon the object, for light shown on the left half,  is the projector associated with the $+1$ eigenvalue of $\hat{O}_i$ with the location of the boundary between the two measurement regions, $x=\beta_i$, being at angstrom resolution. Such spatial resolution has been achieved in optical imaging of single molecules~\cite{Vahid} and in optomechanical experiments~\cite{Ulbricht,Tracy} by using optical interferometry. 

Finally, we note that most of the magnitudes of violations of NSIT or LGI are $\sim 10^{-1}$. Since $n$ number of runs can determine outcome probabilities with uncertainty $\sim 1/\sqrt{n}$, we require $10^4$ experimental runs to ensure that the statistical error is 1 order of magnitude less than the mean value of the violations.
%Also, testing LGI involves determining more number of probabilities compared to testing the NSIT condition. }

{\it Conclusions:--}  We have suitably modified the procedure %used to measure the temporal correlations between variables of a evolving quantum system
for testing LGI and NSIT in order	to show the violation of the classical notion of MR in a manner which is {\em independent} of the parameters: mass, momentum, and frequency. Moreover, this modification offers a quantum jump in simplifying experimental efforts  in terms of measurement precision (even a more coarse-grained measurement than the SQL is sufficient) and tuning of the parameters. Naturally,  this enormously broadens the scope for evidencing nonclassicality for large masses.  No nonclassical state,  such as a quantum superposition of distinct states (e.g., a Schr\"{o}dinger cat state) or even a squeezed state, needs to be prepared \textit{a priori}. Moreover, this approach does not require coupling with any ancillary quantum system or using nonlinearity. Rather the starting point of our  scheme is the most ``classical-like'' of all quantum states--namely, the coherent state, which has been prepared by feedback cooling in several systems \cite{sub-kelvin-Giesler}, including 10 kg LIGO masses \cite{Chris}, and is imminent in several other systems. In fact, this can be regarded as a \textit{scale-invariant test of nonclassicality}: the experimental data curves of MR violations with $t_2$ at the same fractions of $T$ with widely different masses can be made to coincide with each other by adjusting $p_0$ and $\omega$.  

{\bf Note Added:--} A paper has appeared in parallel \cite{HalliwellLatest} reporting QM violations of the different forms of LGI for coherent states of a harmonic
oscillator, focusing on maximizing these violations. In particular, the observable measured at different instants is taken to be the same while testing LGI. Consequently, mass-independent quantum violation has not been achieved in that paper.

{\it Acknowledgements:--} We acknowledge fruitful discussions with Jonathan Halliwell and  Clement Mawby. DD acknowledges the Royal Society (United Kingdom) for the support through the Newton International Fellowship (No. NIF$\backslash$R$1\backslash212007$).  DH acknowledges support from NASI Senior Scientist Fellowship and QuEST-DST Project No. Q-98 of the Government of India. HU would like to acknowledge support EPSRC through Grants No. EP/W007444/1, No. EP/V035975/1 and No. EP/V000624/1, the Leverhulme Trust (No. RPG-2022-57), the EU Horizon 2020 FET-Open project TeQ (No. 766900), and the EU EIC Pathfinder project QuCoM (No. 10032223). SB would like to acknowledge EPSRC Grant No. EP/N031105/1 and No. EP/S000267/1, and STFC Grant No. ST/W006227/1. SB and HU would like to acknowledge EPSRC Grant No. EP/X009467/1.

	\onecolumngrid
	
	\newpage
	
	\appendix
	
	\section{Deriving the form of two-time Leggett-Garg inequality used in this paper}\label{app1}

	We consider the experimental scenario relevant to the treatment given in this paper where an observable $Q$ is
	measured at the instant $t=t_1$ and another observable $R$ is measured at $t=t_2$, where $t_1 < t_2$.  Here, the measured values of $Q$ and $R$ can be $+ 1$ or $-1$ depending on the state of the system.
	
	Now, `realism' implies that $Q$ and $R$ can each be assigned a value $\pm 1$ at any instant, independent of measurement. Let $v(Q)$ and $v(R)$ be such values assigned to $Q$ at instant $t=t_1$ and $R$ at $t=t_2$ respectively. Since $v(Q) = \pm 1$, $v(R) = \pm 1$, we have for $s_1 = \pm 1$ and $s_2 = \pm 1$ that
	\begin{align}
		\Big(1 + s_1 v(Q) \Big) \Big(1 + s_2 v(R) \Big) = 0 \, \, \text{or}, \, \, 4.
	\end{align}
	
	Hence, the above expression leads to the following,
	\begin{align}
		1 + s_1 \langle Q \rangle_{\text{G}} + s_2 \langle R  \rangle_{\text{G}} + s_1 s_2 \langle QR \rangle_{\text{G}} \geq 0,
		\label{gren2}
	\end{align}
	where $\langle \cdots \rangle_\text{G}$ denotes the average over a grand ensemble, where the two observables
	$Q$ and $R$ are measured at instants $t=t_1$ and $t=t_2$ respectively.
	
	Next, using the assumption of noninvasive measurability, we obtain
	\begin{align}
		\langle R  \rangle_{\overline{Q}} = \langle R  \rangle_{\text{G}},
		\label{suben}
	\end{align}
	where $\langle \cdots \rangle_{\overline{Q}}$ is an average over an ensemble identical to the above-mentioned grand
	ensemble, with the exception that the observable $Q$  is not
	measured.
	
	Hence, from Eqs.(\ref{gren2}) and (\ref{suben}) we get the following modified two-time LGI,
	\begin{align}
		1 + s_1 \langle Q \rangle_{\text{G}} + s_2 \langle R  \rangle_{\overline{Q}} + s_1 s_2 \langle QR \rangle_{\text{G}} \geq 0.
		\label{suben2}
	\end{align}

	\section{Deriving the observable probabilities for the time-evolving Schr\"{o}dinger coherent state of a linear harmonic oscillator}\label{app2}

	Consider the following initial Gaussian wave function
	peaked at $x=0$ at instant $t=0$,
	\begin{align}
		\psi(x,t=0) &= N_s  \text{exp}\left(-\frac{x^2}{4 \sigma_0^2} + \frac{\mathbbm{i} p_0 x}{\hbar}\right) \nonumber \\
		& \text{with} \, \, \, \, \, N_s = \sqrt{\frac{1}{\sqrt{2 \pi} \sigma_0}}
	\end{align}
	with the initial momentum expectation value $p_0$, and the
	width $\sigma_0 = \sqrt{\hbar/(2 m \omega)}$, where $\omega$ is the angular frequency of oscillation and $m$ is the mass.
	
	The state evolved under the linear
	harmonic potential  from instant $t=t_i$ to $t=t_j$ (with $t_j - t_i = \Delta_t$) can be evaluated using the following propagator,
	\begin{align}
		K(x', t = t_i; x, t=t_j) &= N_p \text{exp} \left[ \frac{\mathbbm{i} m \omega}{2 \hbar \sin \omega \Delta_t} \left\{ \left( x^2 + x'^2 \right) \cos \omega \Delta_t - 2 x x' \right\} \right] \nonumber \\
		& \text{with} \, \, \, \, \, N_p = \sqrt{\frac{m \omega}{2 \pi \mathbbm{i} \hbar \sin \omega \Delta_t}}
	\end{align}
	
	At the instant $t=t_1=0$, we perform the 
	dichotomic measurement of the observable $Q$ that determines whether the oscillating particle is in the state $1$ (in between $x=\beta_1$ and $x \rightarrow - \infty$), or in the state $2$ (in between $x=\beta_1$ and $x \rightarrow  +\infty$). Here $\beta_1$ is a real number. The outcomes $-1$ and $+1$ correspond to finding the particle in the state $1$ and in the state $2$ respectively.
	
	At the instant $t=t_2$, we perform another
	dichotomic measurement of the observable $R$  that determines whether the oscillating particle is in the state $3$ (in between $x=\beta_2$ and $x \rightarrow - \infty$), or in the state $4$ (in between $x=\beta_2$ and $x \rightarrow  +\infty$). Here,  $\beta_2$ is a real number. The outcomes $-1$ and $+1$ correspond to finding the particle in the state $3$ and in the state $4$ respectively.
	
	Let us choose $\beta_1=0$. Depending on the outcomes $-1$ and $+1$ of the above-mentioned measurement 
	at the instant $t=t_1=0$, the post-measurement states (not normalized) are given by,
	\begin{align}
		&|\psi(t=0)\rangle_{-}= \int_{- \infty}^{0} dx \, \psi(x, t=0) |x\rangle, \nonumber \\
		&|\psi(t=0)\rangle_{+}= \int_{0}^{\infty} dx \, \psi(x, t=0) |x\rangle, 
		\label{pms}
	\end{align}

	Hence, the time-evolved unnormalized wave function at the instant $t=t_2$, when the outcome $-1$ is obtained at $t=0$, is given by,
	\begin{align}
		\psi_{-}(x, t=t_2) &= \int_{-\infty}^{0}  K(x', t = 0; x, t = t_2) \, \psi(x',t=0) \, dx'\nonumber \\
		& = N_s N_p \text{exp} \left(\frac{\mathbbm{i} m \omega x^2 \cos (\omega t_2)}{2 \hbar \sin (\omega t_2)}\right) \int_{-\infty}^{0} dx' \text{exp} \left( - a x'^2 + \mathbbm{i} b x' \right)
	\end{align}
	with
	\begin{align}
		&a =  \frac{1}{4 \sigma_0^2} - \frac{\mathbbm{i} m \omega \cos (\omega t_2)}{2 \hbar \sin(\omega t_2)}, \nonumber \\
		&b   = \frac{p_0}{\hbar} - \frac{m \omega x}{\hbar \sin (\omega t_2)}. \nonumber
	\end{align}
	
	Similarly, the time-evolved unnormalized wave function at the instant $t=t_2$, when the outcome $+1$ is obtained at $t=0$, is given by,
	\begin{align}
		\psi_{+}(x, t=t_2) &= \int_{0}^{\infty}  K(x', t = 0; x, t = t_2) \, \psi(x',t=0) \, dx'\nonumber \\
		& = N_s N_p \text{exp} \left(\frac{\mathbbm{i} m \omega x^2 \cos (\omega t_2)}{2 \hbar \sin (\omega t_2)}\right) \int_{0}^{\infty} dx' \text{exp} \left( - a x'^2 + \mathbbm{i} b x' \right)
	\end{align}
	
	Now, it can be checked that
	\begin{align}
		&\int_{-\infty}^{0} dx' \text{exp} \left( - a x'^2 + \mathbbm{i} b x' \right) = \frac{\sqrt{\pi}}{2\sqrt{a}}\text{exp}\left(- \frac{b^2}{4 a} \right) \left[ 1 - \text{erf} \left( \frac{\mathbbm{i} b}{2 \sqrt{a}}\right) \right], \nonumber \\
		&\int_{0}^{\infty} dx' \text{exp} \left( - a x'^2 + \mathbbm{i} b x' \right) = \frac{\sqrt{\pi}}{2\sqrt{a}}\text{exp}\left(- \frac{b^2}{4 a} \right) \left[ 1 + \text{erf} \left( \frac{\mathbbm{i} b}{2 \sqrt{a}}\right) \right],
	\end{align}
	where the error function $\text{erf}(z) = (2/\sqrt{\pi}) \int_{0}^{z} dt \, \text{exp}(-t^2)$.
	
	Hence, we have 
	\begin{align}
		\psi_{\pm}(x, t=t_2)  = N_s N_p \text{exp} \left(\frac{\mathbbm{i} m \omega x^2 \cos (\omega t_2)}{2 \hbar \sin (\omega t_2)}\right) \frac{\sqrt{\pi}}{2\sqrt{a}}\text{exp}\left(- \frac{b^2}{4 a} \right) \left[ 1 \pm \text{erf} \left( \frac{\mathbbm{i} b}{2 \sqrt{a}}\right) \right].
	\end{align}
	
	We, therefore, have the following forms of joint probabilities
	\begin{align}
		p(Q\pm, R-) &= \left| N_s \right|^2 \left| N_p \right|^2 \frac{\pi}{4 |a|} \int_{-\infty}^{\beta_2} dx \left|\text{exp}\left( -\frac{b^2}{4 a} \right) \right|^2  \left| 1 \pm \text{erf} \left( \frac{\mathbbm{i} b}{2 \sqrt{a}}\right) \right|^2, \nonumber \\
		&= \frac{\sqrt{\frac{m \omega}{\hbar}}}{4 \sqrt{\pi}} \int_{-\infty}^{\beta_2} dx \, \text{exp}\left[ - \left( x \sqrt{\frac{m \omega}{\hbar}} - \frac{p_0 \sin (\omega t_2)}{\sqrt{\hbar m \omega }}  \right)^2 \right] \left| 1 \pm \text{erf}\left[ \frac{\mathbbm{i} \left( \frac{p_0 \sin (\omega t_2)}{\sqrt{\hbar m \omega }} - x \sqrt{\frac{m \omega}{\hbar}} \right)}{\sin (\omega t_2) \sqrt{2 - 2 \mathbbm{i} \cot (\omega t_2) }} \right] \right|^2,
	\end{align}
	and
	\begin{align}
		p(Q\pm, R+) &= \left| N_s \right|^2 \left| N_p \right|^2 \frac{\pi}{4 |a|} \int_{\beta_2}^{\infty} dx \left|\text{exp}\left(- \frac{b^2}{4 a} \right) \right|^2  \left| 1 \pm \text{erf} \left( \frac{\mathbbm{i} b}{2 \sqrt{a}}\right) \right|^2, \nonumber \\
		&= \frac{\sqrt{\frac{m \omega}{\hbar}}}{4 \sqrt{\pi}} \int_{\beta_2}^{\infty} dx \, \text{exp}\left[ - \left( x \sqrt{\frac{m \omega}{\hbar}} - \frac{p_0 \sin (\omega t_2)}{\sqrt{\hbar m \omega }}  \right)^2 \right] \left| 1 \pm \text{erf}\left[ \frac{\mathbbm{i} \left( \frac{p_0 \sin (\omega t_2)}{\sqrt{\hbar m \omega }} - x \sqrt{\frac{m \omega}{\hbar}} \right)}{\sin (\omega t_2) \sqrt{2 - 2 \mathbbm{i} \cot (\omega t_2) }} \right] \right|^2.
		\label{jointpr0}
	\end{align}
	Now, we perform the following change of variable: $y=x \sqrt{\frac{m \omega}{\hbar}} - \frac{p_0 \sin (\omega t_2)}{\sqrt{\hbar m \omega }}$. With this, the expressions for the above-mentioned joint probabilities become
	\begin{align}
		p(Q\pm, R-) &= \frac{1}{4 \sqrt{\pi}} \int_{-\infty}^{\gamma} dy \, \text{exp}\left[ - y^2 \right] \left| 1 \pm \text{erf}\left[ \frac{ -\mathbbm{i} y}{\sin (\omega t_2) \sqrt{2 - 2 \mathbbm{i} \cot (\omega t_2) }} \right] \right|^2. \nonumber \\
		p(Q\pm, R+) &= \frac{1}{4 \sqrt{\pi}} \int_{\gamma}^{\infty} dy \, \text{exp}\left[ - y^2 \right] \left| 1 \pm \text{erf}\left[ \frac{ -\mathbbm{i} y}{\sin (\omega t_2) \sqrt{2 - 2 \mathbbm{i} \cot (\omega t_2) }} \right] \right|^2,  \hspace{0.5cm} \text{with} \, \, \gamma = \beta_2 \sqrt{\frac{m \omega}{\hbar}} - \frac{p_0 \sin (\omega t_2)}{\sqrt{\hbar m \omega }}.
		\label{jointpr}
	\end{align}
	
	Now, let us evaluate the probabilities $p(Q \pm)$. It can easily be checked that 
	\begin{align}
		p(Q -) &= \int_{-\infty}^{0} | \psi(x, t=0) |^2= \frac{1}{2},
	\end{align}
	and 
	\begin{align}
		p(Q +) &= \int_{0}^{\infty} | \psi(x, t=0) |^2 = \frac{1}{2}.
	\end{align}
	
	Next, let us evaluate the probabilities $p(R \pm)$. For this, consider that no measurement is performed at $t=t_1 =0$. Under this condition, the time-evolved wave function at the instant $t=t_2$ is given by,
	\begin{align}
		\psi(x, t=t_2) &= \int_{-\infty}^{\infty}  K(x', t = 0; x, t = t_2) \, \psi(x',t=0) \, dx'\nonumber \\
		& = N_s N_p \text{exp} \left(\frac{\mathbbm{i} m \omega x^2 \cos (\omega t_2)}{2 \hbar \sin (\omega t_2)}\right) \int_{-\infty}^{\infty} dx' \text{exp} \left( - a x'^2 + \mathbbm{i} b x' \right)
	\end{align}
	with
	\begin{align}
		&a =  \frac{1}{4 \sigma_0^2} - \frac{\mathbbm{i} m \omega \cos (\omega t_2)}{2 \hbar \sin(\omega t_2)}, \nonumber \\
		&b   = \frac{p_0}{\hbar} - \frac{m \omega x}{\hbar \sin (\omega t_2)}. \nonumber
	\end{align}
	
	It can be checked that
	\begin{align}
		&\int_{-\infty}^{\infty} dx' \text{exp} \left( - a x'^2 + \mathbbm{i} b x' \right) = \frac{\sqrt{\pi}}{\sqrt{a}}\text{exp}\left(- \frac{b^2}{4 a} \right).
	\end{align}
	Using the above integral, we have 
	\begin{align}
		\psi(x, t=t_2)  = N_s N_p \text{exp} \left(\frac{\mathbbm{i} m \omega x^2 \cos (\omega t_2)}{2 \hbar \sin (\omega t_2)}\right) \frac{\sqrt{\pi}}{\sqrt{a}}\text{exp}\left(- \frac{b^2}{4 a} \right).
	\end{align}
	
	Hence, we have the following forms of probabilities
	\begin{align}
		P(R-) &= \left| N_s \right|^2 \left| N_p \right|^2 \frac{\pi}{|a|} \int_{-\infty}^{\beta_2} dx \left|\text{exp}\left(- \frac{b^2}{4 a} \right) \right|^2 \nonumber \\
		&=\frac{1}{2}\left[1+\text{erf}\left(\frac{\beta_2 m \omega - p_0 \sin (\omega t_2)}{\sqrt{\hbar m \omega }}\right) \right],\nonumber \\
		&=\frac{1}{2}\left[1+\text{erf}\left(\gamma\right) \right], \nonumber \\
		P(R+) &= \left| N_s \right|^2 \left| N_p \right|^2 \frac{\pi}{2 |a|} \int_{\beta_2}^{\infty} dx \left|\text{exp}\left(- \frac{b^2}{4 a} \right) \right|^2\nonumber \\
		&=\frac{1}{2}\left[1-\text{erf}\left(\frac{\beta_2 m \omega - p_0 \sin (\omega t_2)}{\sqrt{\hbar m \omega }}\right) \right],\nonumber \\
		&=\frac{1}{2}\left[1-\text{erf}\left(\gamma\right) \right],  \hspace{1cm} \text{with} \, \, \gamma = \beta_2 \sqrt{\frac{m \omega}{\hbar}} - \frac{p_0 \sin (\omega t_2)}{\sqrt{\hbar m \omega }}.
		\label{singlepr}
	\end{align}
	
	%Hence, quantum violation of the NSIT condition given by (\ref{nsit}) can be evaluated using the expressions (\ref{jointpr}) and (\ref{singlepr}) as follows:
	%\begin{align}
	%	J_{-} &= \frac{1}{2}\left[1+\text{erf}\left(\gamma\right) \right] - \frac{1}{4 \sqrt{\pi}} \int_{-\infty}^{\gamma} dy \, \text{exp}\left[ - y^2 \right] \Bigg\{  \left| 1 + \text{erf}\left[ \frac{ -i y}{\sin (\omega t_2) \sqrt{2 - 2 i \cot (\omega t_2) }} \right] \right|^2 \nonumber \\
	%	& \hspace{8cm} +  \left| 1 - \text{erf}\left[ \frac{ -i y}{\sin (\omega t_2) \sqrt{2 - 2 i \cot (\omega t_2) }} \right] \right|^2 \Bigg\}, \nonumber \\
	%J_{+} &= \frac{1}{2}\left[1-\text{erf}\left(\gamma\right) \right] - \frac{1}{4 \sqrt{\pi}} \int_{\gamma}^{\infty} dy \, \text{exp}\left[ - y^2 \right] \Bigg\{  \left| 1 + \text{erf}\left[ \frac{ -i y}{\sin (\omega t_2) \sqrt{2 - 2 i \cot (\omega t_2) }} \right] \right|^2 \nonumber \\
	%& \hspace{8cm} +  \left| 1 - \text{erf}\left[ \frac{ -i y}{\sin (\omega t_2) \sqrt{2 - 2 i \cot (\omega t_2) }} \right] \right|^2 \Bigg\} \nonumber \\
	%&\hspace{6cm} \text{with} \, \, \gamma = \beta \sqrt{\frac{m \omega}{\hbar}} - \frac{p_0 \sin (\omega t_2)}{\sqrt{\hbar m \omega }}.
	%\end{align}
	
	\section{Choosing the boundary $x=\beta_2$ between two regions in case of the second measurement at $t=t_2$}\label{app3}
	
	If the QM description of the dynamical evolution of the coherent state of linear harmonic oscillator satisfies the assumptions of MR (predefined values of $Q$ are revealed by measurements at the initial instant $t=t_1$ which do not affect the subsequent time evolution of the state), then the following condition must hold good: The QM position probability density at the subsequent instant $t=t_2$ without any measurement performed at $t=t_1$ needs to be equal to the weighted sum of  the time evolved probability densities at $t=t_2$ corresponding to the respective outcomes $\pm 1$ obtained by measuring $Q$ at $t=t_1$, mathematically expressed as follows, %probability density at instant $t=t_2$ without any measurement at $t=t_1$ should be equal to the statistical average of (1) probability density at instant $t=t_2$ of those set of preparations for which the value $+1$ is assigned for the observable $Q$ at instant $t=t_1$, and (2) probability density at instant $t=t_2$ of those  preparations for which the value $-1$ is assigned for $Q$ at $t=t_1$. Macrorealism also implies that these pre-defined values of $Q$ can be revealed by preforming measurements at $t=t_1$. Mathematically, these can be expressed as
	\begin{equation}
		|\psi(x,t=t_2)|^2 = p_+ |\psi_{+}(x,t=t_2)|^2 + p_-|\psi_{-}(x,t=t_2)|^2,
		\label{mr}
	\end{equation}
	where $\psi(x,t=t_2)$ is the wave function at $t=t_2$ without any measurement at $t=t_1$; $\psi_{\pm}(x,t=t_2)$ are the normalized wave functions at $t=t_2$ when the outcomes $\pm 1$ are obtained by performing the measurement of $Q$ at $t=t_1$; $p_{+}$ and $p_-$ are the probabilities with which the values $+1$ and $-1$, respectively, are assigned to the observable $Q$ at $t=t_1$. Also, MR implies $p_{\pm}$ should be equal to the observable probabilities $p(Q\pm) = 1/2$. Any deviation from the condition (\ref{mr}) will signify quantum violation of MR.

	\begin{figure}
		{\color{white}
			\centering
			\includegraphics[width=350px,height=200px]{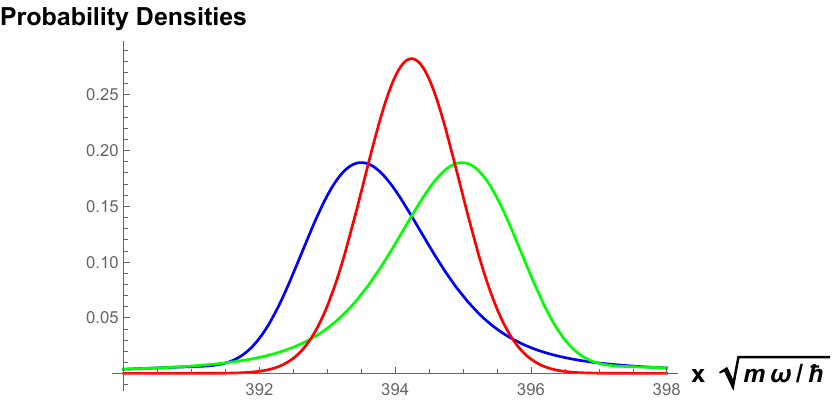}}
		\caption{The blue, green and red curves denote the plots of $|\psi_{-}(x,t=t_2)|^2$, $|\psi_{+}(x,t=t_2)|^2$, $|\psi(x,t=t_2)|^2$ versus $\tilde{x} = x\,\sqrt{\frac{m \omega}{\hbar}}$ respectively for $m=10^{20}$ amu, $p_0 = 3.3\times 10^{-15}$ kg m/s, $\omega = 2\times 10^6$ Hz, $t_2= T/8$. Here we have shown the plots near $\tilde{x}^{(t_2)}_0 = x^{(t_2)}_0\,\sqrt{\frac{m \omega}{\hbar}}$ with $x^{(t_2)}_0 = p_0 \sin (\omega t_2)/(m \omega)$. In other regions, $|\psi_{-}(x,t=t_2)|^2$, $|\psi_{+}(x,t=t_2)|^2$, $|\psi(z,t=t_2)|^2$ become vanishingly small.}
		\label{fig1} 
	\end{figure}

	Next, let us investigate the nature of $|\psi(x,t=t_2)|^2$ and $|\psi_{\pm}(x,t=t_2)|^2$ in the present context to assess a priori  the location  of the boundary for the second measurement (at $x=\beta_2$) most favorable for showing QM violation of Eq.(\ref{mr}). 
	
	First, note that the expression of $|\psi(x,t=t_2)|^2$ is given by,
	\begin{align}
		|\psi(x,t=t_2)|^2 = \frac{\sqrt{\frac{m \omega}{\hbar}}}{\sqrt{\pi}} \text{exp}\left[ - \left( \frac{x - \frac{p_0 \sin (\omega t_2)}{m \omega } }{\sqrt{\frac{\hbar}{m \omega}}} \right)^2 \right].
	\end{align}
	Hence, this is a Gaussian distribution with peak at $x^{(t_2)}_0 = p_0 \sin (\omega t_2)/(m \omega)$ and standard deviation $\Delta^{(t_2)} = \sqrt{\hbar / (2 m \omega)}$.
	
	Next, the expressions of $|\psi_{\pm}(x,t=t_2)|^2$ are given by, 
	\begin{align}
		|\psi_{\pm}(x,t=t_2)|^2 = \frac{\sqrt{\frac{m \omega}{\hbar}}}{4 \sqrt{\pi}} \text{exp}\left[ - \left( \frac{x - \frac{p_0 \sin (\omega t_2)}{m \omega } }{\sqrt{\frac{\hbar}{m \omega}}} \right)^2 \right] \left| 1 \pm \text{erf}\left[ \frac{ - \mathbbm{i} \left( x - \frac{p_0 \sin (\omega t_2)}{m \omega } \right)}{ \sqrt{\frac{\hbar}{m \omega}} \sin (\omega t_2) \sqrt{2 - 2 \mathbbm{i} \cot (\omega t_2) }} \right] \right|^2.
		\label{psimodpm}
	\end{align}
	Now, for some choices of $t_2$, these are Gaussian-like distributions with the peaks being slightly shifted from $x^{(t_2)}_0 = p_0 \sin (\omega t_2)/(m \omega)$. For example, let us take $m=10^{20}$ amu, $p_0 = 3.3\times 10^{-15}$ kg m/s, $\omega = 2\times 10^6$ Hz, $t_2= T/8$, where $T=2 \pi/\omega$ is the time-period. For this choice of the parameters, the plots of $|\psi_{\pm}(x,t=t_2)|^2$ and $|\psi(x,t=t_2)|^2$ are presented in Fig. \ref{fig1}.

\begin{figure}
		{\color{white}
			\centering
			\includegraphics[width=350px,height=200px]{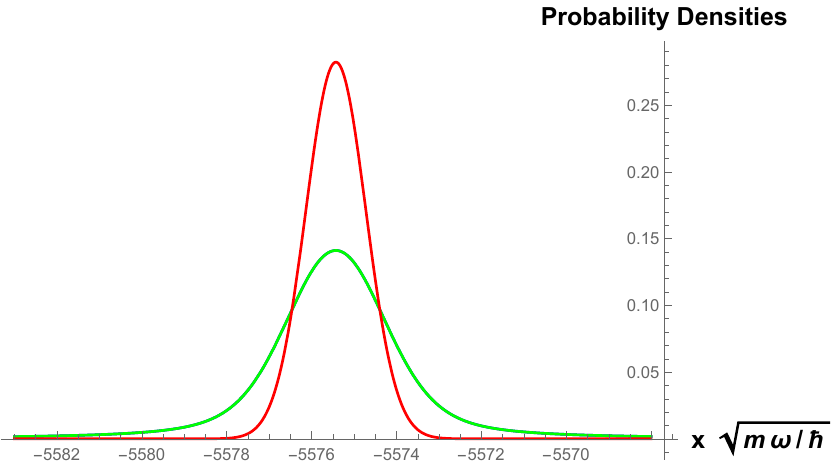}}
		\caption{The green curve denotes the plot of both $|\psi_{-}(x,t=t_2)|^2$ and $|\psi_{+}(x,t=t_2)|^2$ versus $\tilde{x}= x\,\sqrt{\frac{m \omega}{\hbar}}$. On the other hand, the red curve denotes the plot of  $|\psi(x,t=t_2)|^2$ versus $\tilde{x}= x\,\sqrt{\frac{m \omega}{\hbar}}$ for $m=10^{6}$ amu, $p_0 = 3.3\times 10^{-21}$ kg m/s, $\omega = 2\times 10^6$ Hz, $t_2=3 T/4$. Here we have shown the plots near $\tilde{x}_0^{(t_2)} = x^{(t_2)}_0\,\sqrt{\frac{m \omega}{\hbar}}$ with $x^{(t_2)}_0 = p_0 \sin (\omega t_2)/(m \omega)$. In other regions, $|\psi_{-}(x,t=t_2)|^2$, $|\psi_{+}(x,t=t_2)|^2$, $|\psi(z,t=t_2)|^2$ become vanishingly small.}
		\label{fig2} 
	\end{figure}

	Also, for some choices of $t_2$, $|\psi_{\pm}(x,t=t_2)|^2$ given by Eq.(\ref{psimodpm}) are Gaussian-like distributions with the peaks being situated at $x^{(t_2)}_0 = p_0 \sin (\omega t_2)/(m \omega)$, but slightly broadened compared to $|\psi(x,t=t_2)|^2$. For example, let us take $m=10^{6}$ amu, $p_0 = 3.3\times 10^{-21}$ kg m/s, $\omega = 2\times 10^6$ Hz, $t_2=3 T/4$, where $T=2 \pi/\omega$ is the time-period. For this choice of the parameters, the plots of $|\psi(x,t=t_2)|^2$ and $|\psi_{\pm}x,t=t_2)|^2$ are presented in Fig. \ref{fig2}.

	The plots of $|\psi_{\pm}(x,t=t_2)|^2$ and $|\psi(x,t=t_2)|^2$ are of similar natures for other different  choices of parameters.

	From these two plots, we observe the following features:
	
	\begin{itemize}
		\item All the distributions $|\psi(x,t=t_2)|^2$ and $|\psi_{\pm}(x,t=t_2)|^2$ become vanishingly small at $x$ far away from $x^{(t_2)}_0$. 
		
		\item The difference between $|\psi(x,t=t_2)|^2$ and $(1/2)(|\psi_{+}(x,t=t_2)|^2 + |\psi_{-}(x,t=t_2)|^2)$ can be seen at $x=x^{(t_2)}_0 \pm  c \Delta^{(t_2)}$ with $c$ being a  positive number, and $c$ being of the order of $10^{-1}$ or $1$.
	\end{itemize}	
	
	Before proceeding further, it is useful to note that the measurement statistics of $R$ at $t=t_2$ is determined from the area under the curve $|\psi(x,t=t_2)|^2$ or $|\psi_{+}(x,t=t_2)|^2 $ or $|\psi_{-}(x,t=t_2)|^2$ in the region from $x=\beta_2$ to $x=\infty$ or from $x=-\infty$ to $x=\beta_2$. In particular, $p(R +)$ and $p(R -)$ are nothing but the areas under the curve $|\psi(x,t=t_2)|^2$ in the region from $x=\beta_2$ to $x=\infty$ and from $x=-\infty$ to $x=\beta_2$ respectively. On the other hand, $p(Q \pm, R +)$ and $p(Q \pm, R -)$ are the areas under the curve $|\psi_{\pm}(x,t=t_2)|^2$ in the region from $x=\beta_2$ to $x=\infty$ and from $x=-\infty$ to $x=\beta_2$ respectively.
	
	Now, it is seen from Figs. \ref{fig1} and \ref{fig2} that the difference between the areas under the curves $|\psi(x,t=t_2)|^2$ and $(1/2)(|\psi_{+}(x,t=t_2)|^2 + |\psi_{-}(x,t=t_2)|^2)$ in any of the two measurement regions (either from $x=\beta_2$ to $x=\infty$ or from $x=-\infty$ to $x=\beta_2$) vanishes if $\beta_2$ is chosen far away from the region $x^{(t_2)}_0 \pm c \Delta^{(t_2)}$ with $c>0$ being of the order of $10^{-1}$ or $1$. There will be difference between the areas under the curves- $|\psi(x,t=t_2)|^2$ and $(1/2)(|\psi_{+}(x,t=t_2)|^2 + |\psi_{-}(x,t=t_2)|^2)$ in any of the aforementioned two regions only if $\beta_2$ is chosen at $x=x^{(t_2)}_0 \pm c \Delta^{(t_2)}$. Also, if $\beta_2$ is chosen at $x^{(t_2)}_0$, then there is no difference between the areas under the curves- $|\psi(x,t=t_2)|^2$ and $(1/2)(|\psi_{+}(x,t=t_2)|^2 + |\psi_{-}(x,t=t_2)|^2)$ in the region from $x=\beta_2$ to $x=\infty$ or in the region from $x=-\infty$ to $x=\beta_2$. Hence, the difference between $|\psi(x,t=t_2)|^2$ and $(1/2)(|\psi_{+}(x,t=t_2)|^2 + |\psi_{-}(x,t=t_2)|^2)$ is likely to be reflected in the measurement statistics of $Q$ and $R$ if $\beta_2$ is chosen at $x=x^{(t_2)}_0 \pm c \Delta^{(t_2)}$.

	Based on the above consideration, for the purpose of our analysis, we have therefore chosen the boundary between the two measurement regions for the second measurement at $t=t_2$ to be located at $x= \beta_2 = x^{(t_2)}_0 \pm c \Delta^{(t_2)}$ with $c>0$ being of the order of $10^{-1}$ or $1$.

	\section{Details on the practical challenges with large mass}\label{app4}
	
	In practical situations, it is almost impossible to keep the boundary between the two regions (illuminated and unilluminated) to be fixed in all experimental runs. Rather, we can expect that the aforementioned boundary at $t=t_i$  will be at $\beta_i + \epsilon_i$ (where $\epsilon_i$ is a small positive/negative number), where $\epsilon_i$ will be different in different runs. In effect, the observed violation of the two-time NSIT will be the statistical average of different (positive or negative) values of $\text{N}_+$,  $\text{N}_-$ corresponding to different $\epsilon_1$ and $\epsilon_2$. Similarly, the observed violation of the two-time LGI will be the statistical average of different (positive or negative) values of  $\text{L}_{+,+}$, $\text{L}_{+,-}$, $\text{L}_{-,+}$, $\text{L}_{-,-}$ corresponding to  different $\epsilon_1$ and $\epsilon_2$.
	
	Let us, at first, determine the quantum violation of two-time NSIT/LGI with the boundary between the two regions in case of first measurement being located at $x=\beta_1 + \epsilon_1$ and that in case of second measurement being located at $x=\beta_2 + \epsilon_2$, where $\epsilon_1$ and $\epsilon_2$ are two fixed positive or negative numbers.  
	
	The first measurement of $Q$ at $t=t_1 = 0$ determines whether the particle is in the region from $x=-\infty$ to $x = \beta_1 + \epsilon_1$ (corresponds to the outcome $-1$), or  from $x= \beta_1 + \epsilon_1$ to $x=\infty$ (corresponds to the outcome $+1$), where $\beta_1 = 0$ and $\epsilon_1$ can be positive or negative. Depending on the outcomes $-1$ and $+1$ of this measurement, the post-measurement states (not normalized) are given by,
	\begin{align}
		&|\psi(t=0)\rangle_{-}= \int_{- \infty}^{\epsilon_1} dx \, \psi(x, t=0) |x\rangle, \nonumber \\
		&|\psi(t=0)\rangle_{+}= \int_{\epsilon_1}^{\infty} dx \, \psi(x, t=0) |x\rangle, 
		\label{pms22}
	\end{align}

	Hence, the time-evolved unnormalized wave function at the instant $t=t_2$, when the outcome $-1$ is obtained at $t=0$, is given by,
	\begin{align}
		\psi_{-}(x, t=t_2) &= \int_{-\infty}^{\epsilon_1}  K(x', t = 0; x, t = t_2) \, \psi(x',t=0) \, dx'\nonumber \\
		& = N_s N_p \text{exp} \left(\frac{\mathbbm{i} m \omega x^2 \cos (\omega t_2)}{2 \hbar \sin (\omega t_2)}\right) \int_{-\infty}^{\epsilon_1} dx' \text{exp} \left( - a x'^2 + \mathbbm{i} b x' \right)
	\end{align}
	with
	\begin{align}
		&a =  \frac{1}{4 \sigma_0^2} - \frac{\mathbbm{i} m \omega \cos (\omega t_2)}{2 \hbar \sin(\omega t_2)}, \nonumber \\
		&b   = \frac{p_0}{\hbar} - \frac{m \omega x}{\hbar \sin (\omega t_2)}. \nonumber
	\end{align}
	
	Similarly, the time-evolved unnormalized wave function at the instant $t=t_2$, when the outcome $+1$ is obtained at $t=0$, is given by,
	\begin{align}
		\psi_{+}(x, t=t_2) &= \int_{\epsilon_1}^{\infty}  K(x', t = 0; x, t = t_2) \, \psi(x',t=0) \, dx'\nonumber \\
		& = N_s N_p \text{exp} \left(\frac{\mathbbm{i} m \omega x^2 \cos (\omega t_2)}{2 \hbar \sin (\omega t_2)}\right) \int_{\epsilon_1}^{\infty} dx' \text{exp} \left( - a x'^2 + \mathbbm{i} b x' \right)
	\end{align}
	
	It can be checked that
	\begin{align}
		&\int_{-\infty}^{\epsilon_1} dx' \text{exp} \left( - a x'^2 + \mathbbm{i} b x' \right) = \frac{\sqrt{\pi}}{2\sqrt{a}}\text{exp}\left(- \frac{b^2}{4 a} \right) \left[ 1 - \text{erf} \left(  \frac{\mathbbm{i} \left( b + 2 \mathbbm{i} a \epsilon_1 \right) }{2 \sqrt{a}}\right) \right], \nonumber \\
		&\int_{\epsilon_1}^{\infty} dx' \text{exp} \left( - a x'^2 + \mathbbm{i} b x' \right) = \frac{\sqrt{\pi}}{2\sqrt{a}}\text{exp}\left(- \frac{b^2}{4 a} \right) \left[ 1 + \text{erf} \left( \frac{\mathbbm{i} \left( b + 2 \mathbbm{i} a \epsilon_1 \right) }{2 \sqrt{a}}\right) \right].
	\end{align}

	We, therefore, have the following,
	\begin{align}
		\psi_{\pm}(x, t=t_2)  = N_s N_p \text{exp} \left(\frac{\mathbbm{i} m \omega x^2 \cos (\omega t_2)}{2 \hbar \sin (\omega t_2)}\right) \frac{\sqrt{\pi}}{2\sqrt{a}}\text{exp}\left(- \frac{b^2}{4 a} \right) \left[ 1 \pm \text{erf} \left( \frac{\mathbbm{i} \left( b + 2 \mathbbm{i} a \epsilon_1 \right) }{2 \sqrt{a}}\right) \right].
	\end{align}
	
	Next, let us also consider that the second measurement of $R$  at $t=t_2$ determines whether the particle is in the region from $x=-\infty$ to $x = \beta_2 + \epsilon_2$ (corresponds to the outcome $-1$), or  from $x= \beta_2 + \epsilon_2$ to $x=\infty$ (corresponds to the outcome $+1$), where $\beta_2 = p_0 \sin (\omega t_2)/(m \omega ) + \sqrt{\hbar /( m \omega)}$ as mentioned in the main paper and $\epsilon_2$ is a real number. Hence, we have the following expressions of joint probabilities,
	\begin{align}
		p(Q\pm, R-) &= \left| N_s \right|^2 \left| N_p \right|^2 \frac{\pi}{4 |a|} \int_{-\infty}^{\beta_2+\epsilon_2} dx \left|\text{exp}\left( -\frac{b^2}{4 a} \right) \right|^2  \left| 1 \pm \text{erf} \left( \frac{ \mathbbm{i} \left(  b + 2 \mathbbm{i} a \epsilon_1 \right)}{2 \sqrt{a}}\right) \right|^2, \nonumber \\
		&= \frac{\sqrt{\frac{m \omega}{\hbar}}}{4 \sqrt{\pi}} \int_{-\infty}^{\beta_2 + \epsilon_2} dx \, \text{exp}\left[ - \left( x \sqrt{\frac{m \omega}{\hbar}} - \frac{p_0 \sin (\omega t_2)}{\sqrt{\hbar m \omega }}  \right)^2 \right] \left| 1 \pm \text{erf}\left[ \frac{\mathbbm{i} \left( \frac{p_0 \sin (\omega t_2)}{\sqrt{\hbar m \omega }} - x \sqrt{\frac{m \omega}{\hbar}} +  \epsilon_1 \sqrt{\frac{m \omega}{\hbar}} \text{exp}\left[ \mathbbm{i} \omega t_2 \right] \right)}{\sin (\omega t_2) \sqrt{2 - 2 \mathbbm{i} \cot (\omega t_2) }} \right] \right|^2,
	\end{align}
	and
	\begin{align}
		p(Q\pm, R+) &= \left| N_s \right|^2 \left| N_p \right|^2 \frac{\pi}{4 |a|} \int_{\beta_2+\epsilon_2}^{\infty} dx \left|\text{exp}\left( -\frac{b^2}{4 a} \right) \right|^2  \left| 1 \pm \text{erf} \left( \frac{ \mathbbm{i}\left( b + 2 \mathbbm{i} a \epsilon_1 \right)}{2 \sqrt{a}}\right) \right|^2, \nonumber \\
		&= \frac{\sqrt{\frac{m \omega}{\hbar}}}{4 \sqrt{\pi}} \int_{\beta_2+\epsilon_2}^{\infty} dx \, \text{exp}\left[ - \left( x \sqrt{\frac{m \omega}{\hbar}} - \frac{p_0 \sin (\omega t_2)}{\sqrt{\hbar m \omega }}  \right)^2 \right] \left| 1 \pm \text{erf}\left[ \frac{\mathbbm{i} \left( \frac{p_0 \sin (\omega t_2)}{\sqrt{\hbar m \omega }} - x \sqrt{\frac{m \omega}{\hbar}} +  \epsilon_1 \sqrt{\frac{m \omega}{\hbar}} \text{exp}\left[ \mathbbm{i} \omega t_2 \right] \right)}{\sin (\omega t_2) \sqrt{2 - 2 \mathbbm{i} \cot (\omega t_2) }} \right] \right|^2.
		\label{jointpr022}
	\end{align}
	Now, we perform the following change of variable: $y=x \sqrt{\frac{m \omega}{\hbar}} - \frac{p_0 \sin (\omega t_2)}{\sqrt{\hbar m \omega }}$. Also, let us define the following
	\begin{align}
		\widetilde{\epsilon}_1 =  \epsilon_1 \sqrt{\frac{m \omega}{\hbar}}, \hspace{0.5cm} \widetilde{\epsilon}_2 =  \epsilon_2 \sqrt{\frac{m \omega}{\hbar}}
	\end{align}
	With these, the expressions for the above-mentioned joint probabilities become
	\begin{align}
		p(Q\pm, R-) &= \frac{1}{4 \sqrt{\pi}} \int_{-\infty}^{1+\widetilde{\epsilon}_2} dy \, \text{exp}\left[ - y^2 \right] \left| 1 \pm \text{erf}\left[ \frac{ \mathbbm{i}\left(-y + \widetilde{\epsilon}_1 \, \text{exp}\left[ \mathbbm{i} \omega t_2 \right] \right) }{\sin (\omega t_2) \sqrt{2 - 2 \mathbbm{i} \cot (\omega t_2) }} \right] \right|^2. \label{appe1} \\
		p(Q\pm, R+) &= \frac{1}{4 \sqrt{\pi}} \int_{1+\widetilde{\epsilon}_2}^{\infty} dy \, \text{exp}\left[ - y^2 \right] \left| 1 \pm \text{erf}\left[ \frac{ \mathbbm{i}\left(-y + \widetilde{\epsilon}_1 \, \text{exp}\left[ \mathbbm{i} \omega t_2 \right] \right) }{\sin (\omega t_2) \sqrt{2 - 2 \mathbbm{i} \cot (\omega t_2) }} \right] \right|^2.
		\label{appe2}
	\end{align}
	
	Now, let us evaluate the probabilities $p(Q \pm)$. It can easily be checked that 
	\begin{align}
		p(Q -) &= \int_{-\infty}^{\epsilon_1} | \psi(x, t=0) |^2= \frac{1}{2} \left[1+ \text{erf}\left(\widetilde{\epsilon}_1 \right)\right], \label{appe3} \\
		p(Q +) &= \int_{\epsilon_1}^{\infty} | \psi(x, t=0) |^2 = \frac{1}{2} \left[1- \text{erf}\left(\widetilde{\epsilon}_1 \right)\right].
		\label{appe4}
	\end{align}
	
	Next, we will evaluate the probabilities $p(R \pm)$. For this, we consider that no measurement is performed at $t=t_1 =0$. With this, the time-evolved wave function at the instant $t=t_2$ is given by,
	\begin{align}
		\psi(x, t=t_2) &= \int_{-\infty}^{\infty}  K(x', t = t_1; x, t = t_2) \, \psi(x',t=t_1) \, dx'\nonumber \\
		& = N_s N_p \text{exp} \left(\frac{\mathbbm{i} m \omega x^2 \cos (\omega t_2)}{2 \hbar \sin (\omega t_2)}\right) \int_{-\infty}^{\infty} dx' \text{exp} \left( - a x'^2 + \mathbbm{i} b x' \right) \nonumber \\
		&= N_s N_p \text{exp} \left(\frac{\mathbbm{i} m \omega x^2 \cos (\omega t_2)}{2 \hbar \sin (\omega t_2)}\right) \frac{\sqrt{\pi}}{\sqrt{a}}\text{exp}\left(- \frac{b^2}{4 a} \right).
	\end{align}
	with
	\begin{align}
		&a =  \frac{1}{4 \sigma_0^2} - \frac{\mathbbm{i} m \omega \cos (\omega t_2)}{2 \hbar \sin(\omega t_2)}, \nonumber \\
		&b   = \frac{p_0}{\hbar} - \frac{m \omega x}{\hbar \sin (\omega t_2)}. \nonumber
	\end{align}

	\begin{figure}
		{\color{white}
			\centering
			\includegraphics[width=350px,height=200px]{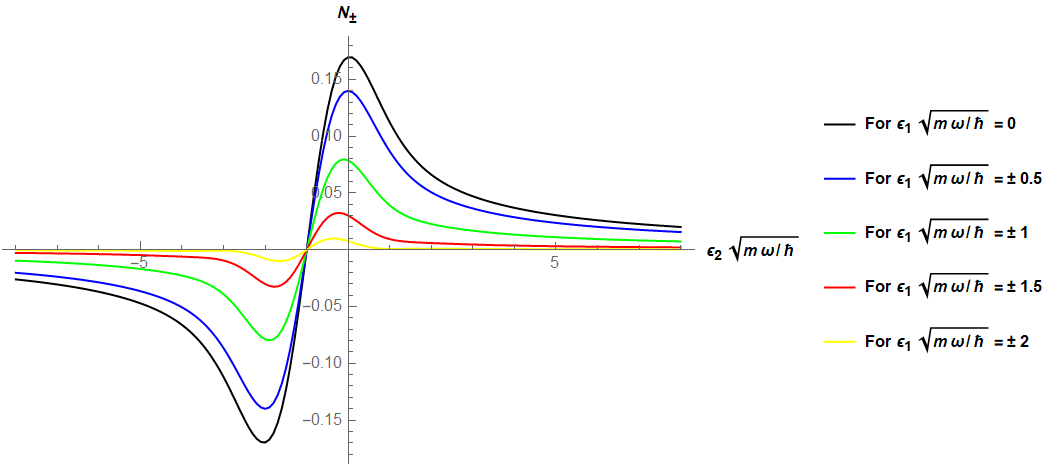}}
		\caption{Plot of quantum violation of the two-time NSIT condition ($\text{N}_{\pm}$) with different choices of $\widetilde{\epsilon}_1 =  \epsilon_1 \sqrt{m \omega/\hbar}$ versus $\widetilde{\epsilon}_2 =  \epsilon_2 \sqrt{m \omega/\hbar}$ for $t_2 = T/4$ and for any choice of $m$, $\omega$, $p_0$.}
		\label{fig3} 
	\end{figure}

	Therefore, we have the following form of probabilities
	\begin{align}
		P(R-) &= \left| N_s \right|^2 \left| N_p \right|^2 \frac{\pi}{|a|} \int_{-\infty}^{\beta_2+\epsilon_2} dx \left|\text{exp}\left(- \frac{b^2}{4 a} \right) \right|^2 \nonumber \\
		&=\frac{1}{2}\left[1+\text{erf}\left(\frac{\beta_2 m \omega + \epsilon_2 m \omega - p_0 \sin (\omega t_2)}{\sqrt{\hbar m \omega }}\right) \right],\nonumber \\
		&=\frac{1}{2}\left[1+\text{erf}\left(1+ \widetilde{\epsilon}_2 \right) \right], \label{appe5} \\
		P(R+) &= \left| N_s \right|^2 \left| N_p \right|^2 \frac{\pi}{2 |a|} \int_{\beta_2 +\epsilon_2}^{\infty} dx \left|\text{exp}\left(- \frac{b^2}{4 a} \right) \right|^2\nonumber \\
		&=\frac{1}{2}\left[1-\text{erf}\left(\frac{\beta_2 m \omega + \epsilon_2 m \omega - p_0 \sin (\omega t_2)}{\sqrt{\hbar m \omega }}\right) \right],\nonumber \\
		&=\frac{1}{2}\left[1-\text{erf}\left(1+\widetilde{\epsilon}_2 \right) \right].
		\label{appe6}
	\end{align}
	
	Using the expressions mentioned in Eqs.(\ref{appe1}), (\ref{appe2}), (\ref{appe3}), (\ref{appe4}), (\ref{appe5}), (\ref{appe6}), the two-time NSIT condition or the two-time LGI can be tested. 
	
	Let us take $t_2 = T/4$ as an example. For this choice of $t_2$, we plot  $\text{N}_{\pm}$ with different choices of $\widetilde{\epsilon}_1 =  \epsilon_1 \sqrt{m \omega/\hbar}$ versus $\widetilde{\epsilon}_2 =  \epsilon_2 \sqrt{m \omega/\hbar}$ for any choice of $m$, $\omega$, $p_0$ in Fig. \ref{fig3}. It is evident from this figure that the permissible ranges of $\epsilon_1$ and $\epsilon_2$ for which one can get quantum violation of the NSIT condition are proportional to $1/(m \omega)$. It can be checked that this feature persists for other choices of $t_2$ as well.
	
	As mentioned earlier, in realistic situation, the boundary between the two regions in case of the measurements at $t=t_i$ will be at $x=\beta_i + \epsilon_i$, where $\epsilon_i$ will be different in different experimental runs. Precise measurement implies that $\epsilon_i$ is zero in each experimental run. On the other hand, with an increase in the imprecision, more non-zero values of $\epsilon_i$ around $0$ will appear in different runs. In other words, as the measurement imprecision increases, the boundaries between the two regions at $t=t_i$ in different runs will be located at more number of different points around $\beta_i$.
	
	Now, the observed quantum violation of the two-time NSIT condition will be the statistical average of  different values of $N_+$ or $N_-$ corresponding to different $\epsilon_1$ and $\epsilon_2$.  Hence, the precision of the two measurements required to get significant quantum violation of the NSIT condition increases with increasing $m$ for any $\omega$. Moreover, for large $m$, this demand for increasing precision can be countered by decreasing $\omega$.
	
	In a similar way, it can be shown that the permissible ranges of $\epsilon_1$ and $\epsilon_2$ for which one can get quantum violation of  the two-time LGI are proportional to $1/(m \omega)$. Hence, in this case also, the precision of the two measurements required to get significant quantum violation of the two-time LGI increases with increasing values of $m$ for any $\omega$. Also, for large $m$, this necessity for increasing precision in measurements can be avoided by decreasing the angular frequency $\omega$.
	
	%of it can be observed that the violation of NSIT is maximum for $\widetilde{\epsilon} =0$. As $\widetilde{\epsilon}$ increases from $0$, the amount of violation decreases and, finally, vanishes. On the other hand, as $\widetilde{\epsilon}$ decreases from $0$, the amount of violation decreases and then after a certain value of $\widetilde{\epsilon}$ the violation becomes negative.

	\section{Measurement imprecision}\label{app5}
	
	Following the calculations presented in Sec. \ref{app4}, it can be shown that probing quantum violation of the NSIT condition for any $m$, $\omega$ and $p_0$ in our proposed setup is possible even when the accuracies of the coarse-grained position measurements are much worse than the standard quantum limit. 
	
\begin{figure}
		{\color{white}
			\centering
			\includegraphics[width=400px,height=200px]{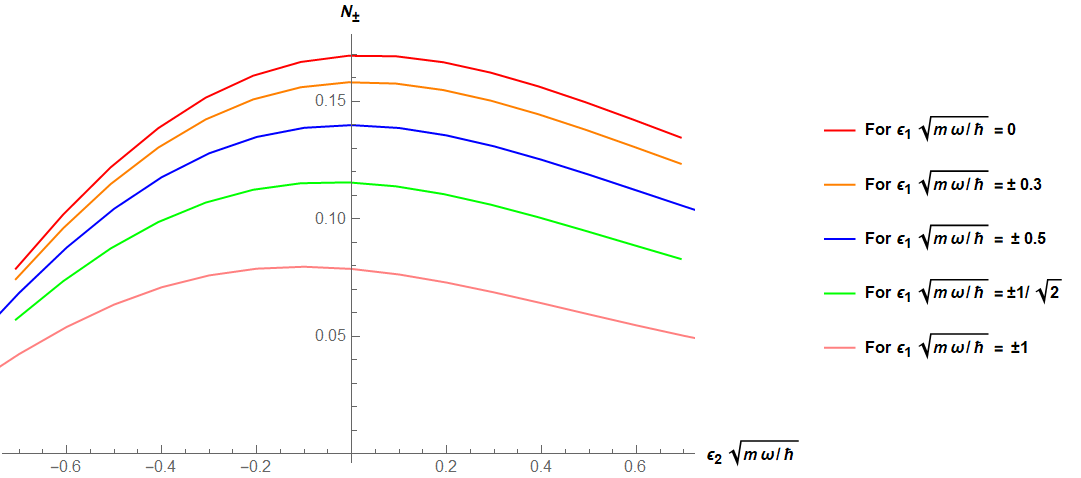}}
		\caption{Plot of quantum violation of the two-time NSIT condition ($\text{N}_{\pm}$) with different choices of $\widetilde{\epsilon}_1 =  \epsilon_1 \sqrt{m \omega/\hbar}$ versus $\widetilde{\epsilon}_2 =  \epsilon_2 \sqrt{m \omega/\hbar}$ for $t_2 = T/4$ and for any choice of $m$, $\omega$, $p_0$.}
		\label{fig4} 
	\end{figure}

	When imprecisions in the coarse-grained position measurements, denoted by $\delta x$, lie in between $\delta x = + \sqrt{\hbar/m \omega}$ and $\delta x = - \sqrt{\hbar/m \omega}$ (which is equivalent to measurement imprecision much worse than the standard quantum limit for the system considered by us), then the observed violation will be the statistical average of different values of $N_{\pm}$ corresponding to different values of $\epsilon_1$ and $\epsilon_2$ with $- \sqrt{\hbar/m \omega} \leq \epsilon_1 \leq  \sqrt{\hbar/m \omega}$ and  $- \sqrt{\hbar/m \omega} \leq \epsilon_2 \leq  \sqrt{\hbar/m \omega}$. From Fig. \ref{fig4}, it is evident that one can observe quantum violation when the measurement imprecision lies in the above-mentioned range. Note that Fig. \ref{fig4} is for  $t_2 = T/4$. For other values of $t_2$ mentioned in the Table 1 of the main paper, one can get similar plots and, hence, same conclusion can be drawn for other choices of $t_2$.
	
	%However, we observe that measurements to the accuracy of the standard quantum limit are not sufficient for showing quantum violation of the two-time LGI. Since quantum violation of any of the two conditions, LGI and NSIT,  is a signature of genuine quantumness, one can use the two-time NSIT condition for probing quantumness in our setup when measurement precision beyond the standard quantum limit is not achievable.
	
\begin{thebibliography}{1}
	
	\bibitem{leggett02} A. J. Leggett, \emph{Testing the limits of quantum mechanics: motivation, state of play, prospects}, \href{https://doi.org/10.1088/0953-8984/14/15/201}{J. Phys. Condens. Matter \textbf{14}, R415 (2002)}.
	
	\bibitem{swfc1} G. C. Ghirardi, A. Rimini, and T. Weber, \emph{Unified dynamics for microscopic and macroscopic systems}, \href{https://doi.org/10.1103/PhysRevD.34.470}{Phys. Rev. D \textbf{34}, 470 (1986)}.
	
	\bibitem{swfc2} P. Pearle, \emph{Combining stochastic dynamical state-vector reduction with spontaneous localization}, \href{https://doi.org/10.1103/PhysRevA.39.2277}{Phys. Rev. A \textbf{39}, 2277 (1989)}.
	
	\bibitem{BassiRMP} A. Bassi, K. Lochan, S. Satin, T. P. Singh, and H. Ulbricht, \emph{Models of wave-function collapse, underlying theories, and experimental tests}, \href{https://doi.org/10.1103/RevModPhys.85.471}{Rev. Mod. Phys. \textbf{85}, 471 (2013)}.
	
	\bibitem{Bose2017} S. Bose, A. Mazumdar, G. W. Morley, H. Ulbricht, M. Toros,  M. Paternostro, A. A. Geraci, P. F. Barker, M. S. Kim, and G. Milburn, \emph{Spin Entanglement Witness for Quantum Gravity}, \href{https://doi.org/10.1103/PhysRevLett.119.240401}{Phys. Rev. Lett. \textbf{119}, 240401 (2017)}.
	
	\bibitem{Marletto2017} C. Marletto, and V. Vedral, \emph{Gravitationally Induced Entanglement between Two Massive Particles is Sufficient Evidence of Quantum Effects in Gravity}, \href{https://doi.org/10.1103/PhysRevLett.119.240402}{Phys. Rev. Lett. \textbf{119}, 240402 (2017)}.
	
	\bibitem{Marshman2020} R. J. Marshman, A. Mazumdar, and S. Bose, \emph{Locality and entanglement in table-top testing of the quantum nature of linearized gravity},
	\href{https://doi.org/10.1103/PhysRevA.101.052110}{Phys. Rev. A \textbf{101}, 052110 (2020)}.	
	
	\bibitem{Bose2022} S. Bose, A. Mazumdar, M. Schut, and M. Toros, \emph{Mechanism for the quantum natured gravitons to entangle masses}, \href{https://doi.org/10.1103/PhysRevD.105.106028}{Phys. Rev. D \textbf{105}, 106028 (2022)}.
	
	\bibitem{Rovelli2019} M. Christodoulou, and C. Rovelli, \emph{On the possibility of laboratory evidence for quantum superposition of geometries}, \href{https://doi.org/10.1016/j.physletb.2019.03.015}{Phys. Lett. B \textbf{792}, 64 (2019)}.
	
	\bibitem{Christodoulou2022} M. Christodoulou, A. D. Biagio, M. Aspelmeyer, C. Brukner, C. Rovelli, and R. Howl, \emph{Locally mediated entanglement through gravity from first principles}, \href{https://arxiv.org/abs/2202.03368}{arXiv:2202.03368 [quant-ph]}.
	
	\bibitem{LHO1} S. Bose, D. Home, and S. Mal, \emph{Nonclassicality of the Harmonic-Oscillator Coherent State Persisting up to the Macroscopic Domain}, \href{https://doi.org/10.1103/PhysRevLett.120.210402}{Phys. Rev. Lett. \textbf{120}, 210402 (2018)}.
	
	\bibitem{LHO2} J. J. Halliwell, A. Bhatnagar, E. Ireland, H. Nadeem, and V. Wimalaweera, \emph{Leggett-Garg tests for macrorealism: Interference experiments and the simple harmonic oscillator}, \href{https://doi.org/10.1103/PhysRevA.103.032218}{Phys. Rev. A \textbf{103}, 032218 (2021)}.
	
	\bibitem{lgi1} A. J. Leggett and A. Garg, \emph{Quantum mechanics versus macroscopic realism: Is the flux there when nobody looks?}, \href{https://doi.org/10.1103/PhysRevLett.54.857}{Phys. Rev. Lett. \textbf{54}, 857 (1985)}.
	
	\bibitem{nsit}	J. Kofler and C. Brukner, \emph{Condition for macroscopic realism beyond the Leggett-Garg inequalities}, \href{https://doi.org/10.1103/PhysRevA.87.052115}{Phys. Rev. A 87, 052115 (2013)}.
	
	\bibitem{lgi2} A. J. Leggett, \emph{Realism and the physical world}, \href{https://doi.org/10.1088/0034-4885/71/2/022001}{Rep. Prog. Phys. \textbf{71}, 022001 (2008)}.

 	\bibitem{explgi2} G. C. Knee, K. Kakuyanagi, M.-C. Yeh, Y. Matsuzaki, H. Toida, H. Yamaguchi, S. Saito, A. J. Leggett, and W. J. Munro, \emph{A strict experimental test of macroscopic realism in a superconducting flux qubit}, \href{https://doi.org/10.1038/ncomms13253}{Nat. Commun. \textbf{7}, 13253 (2016)}.

  	\bibitem{halliwelllginsit} J. J. Halliwell, \emph{Comparing conditions for macrorealism: Leggett-Garg inequalities versus no-signaling in time}, \href{https://doi.org/10.1103/PhysRevA.96.012121}{Phys. Rev. A \textbf{96}, 012121 (2017)}.

   \bibitem{KneePRA} G. C. Knee, M. Marcus, L. D. Smith, and A. Datta, \emph{Subtleties of witnessing quantum coherence in nonisolated systems}, \href{https://doi.org/10.1103/PhysRevA.98.052328}{Phys. Rev. A \textbf{98}, 052328 (2018)}.
	
	
	\bibitem{qlgi1} C. Emary, N. Lambert, and F. Nori, \emph{Leggett–Garg inequalities}, \href{https://doi.org/10.1088/0034-4885/77/1/016001}{Rep. Prog. Phys. \textbf{77},	016001 (2014)}.
	
	\bibitem{explgi1} J. A. Formaggio, D. I. Kaiser, M. M. Murskyj, and T. E. Weiss, \emph{Violation of the Leggett–Garg Inequality in Neutrino Oscillations}, \href{https://doi.org/10.1103/PhysRevLett.117.050402}{Phys. Rev. Lett. \textbf{117}, 050402 (2016)}.
	

	
	\bibitem{explgi3} K. Joarder, D. Saha, D. Home, and U. Sinha, \emph{Loophole-Free Interferometric Test of Macrorealism Using Heralded Single Photons}, \href{https://doi.org/10.1103/PRXQuantum.3.010307}{PRX Quantum \textbf{3}, 010307 (2022)}.
	
	%\bibitem{wlgi} D. Saha, S. Mal, P. K. Panigrahi, and D. Home, \emph{Wigner's form of the Leggett-Garg inequality, the no-signaling-in-time condition, and unsharp measurements}, \href{https://doi.org/10.1103/PhysRevA.91.032117}{Phys. Rev. A {\bf 91}, 032117 (2015)}.
	
	\bibitem{explgi4} C. Robens, W. Alt, D. Meschede, C. Emary, and A. Alberti, \emph{Ideal Negative Measurements in Quantum Walks Disprove Theories Based on Classical Trajectories}, \href{https://doi.org/10.1103/PhysRevX.5.011003}{Phys. Rev. X \textbf{5}, 011003 (2015)}.
	
	\bibitem{explgi5} G. C. Knee, S. Simmons, E. M. Gauger, J. J. Morton, H. Riemann, N. V. Abrosimov, P. Becker, H.-J. Pohl, K. M. Itoh, M. L. Thewalt, G. A. D. Briggs, and S. C.
	Benjamin, \emph{Violation of a Leggett–Garg inequality with
		ideal non-invasive measurements}, \href{https://doi.org/10.1038/ncomms1614}{Nat. Commun. \textbf{3}, 606 (2012)}.
	
	\bibitem{macro1} S. Gerlich, S. Eibenberger, M. Tomandl, S. Nimmrichter, K. Hornberger, P. J. Fagan, J. T\"{u}xen, M. Mayor, and  M. Arndt, \emph{Quantum interference of large organic molecules}, \href{https://doi.org/10.1038/ncomms1263}{Nat Commun \textbf{2}, 263 (2011)}.
	
	\bibitem{macro2} Y. Y. Fein, P. Geyer, P. Zwick, F. Kialka, S. Pedalino, M. Mayor, S. Gerlich, and M. Arndt, \emph{Quantum superposition of molecules beyond 25 kDa}, \href{https://doi.org/10.1038/s41567-019-0663-9}{Nat. Phys. \textbf{15}, 1242 (2019)}.
	
	
	
	
	\bibitem{lg2} J. J. Halliwell, \emph{Leggett-Garg inequalities and no-signaling in time: A quasiprobability approach}, \href{https://doi.org/10.1103/PhysRevA.93.022123}{Phys. Rev. A \textbf{93}, 022123 (2016)}.

 \bibitem{wlgi} D. Saha, S. Mal, P. K. Panigrahi, and D. Home, \emph{Wigner's form of the Leggett-Garg inequality, the no-signaling-in-time condition, and unsharp measurements}, \href{https://doi.org/10.1103/PhysRevA.91.032117}{Phys. Rev. A {\bf 91}, 032117 (2015)}
	
\bibitem{supp} See the Appendix  for the derivation of two-time LGI using two different observables, the derivations of the observable probabilities in the context of the scenario considered by us, justification behind choosing the particular location of $x=\beta_2$, practical challenge with a large mass, required measurement precision.
	
	%  \bibitem{minlg2} J. J. Halliwell, H. Beck, B. K. B. Lee, and S. O'Brien, \emph{Quasiprobability for the arrival-time problem with links to backflow and the Leggett-Garg inequalities}, \href{https://doi.org/10.1103/PhysRevA.99.012124}{Phys. Rev. A \textbf{99}, 012124 (2019)}.
	
	\bibitem{ot} J. Bateman, S. Nimmrichter, K. Hornberger, and H. Ulbricht, \emph{Near-field interferometry of a free-falling nanoparticle from a point-like source}, \href{https://doi.org/10.1038/ncomms5788}{Nat. Commun. {\bf 5}, 4788 (2014)}.
	
	\bibitem{at1} P. Z. G. Fonseca, E. B. Aranas, J. Millen, T. S. Monteiro, and P. F. Barker, \emph{Nonlinear Dynamics and Strong Cavity Cooling of Levitated Nanoparticles}, \href{https://doi.org/10.1103/PhysRevLett.117.173602}{Phys. Rev. Lett. \textbf{117}, 173602 (2016)}.
	
	\bibitem{at2} J. Millen, P. Z. G. Fonseca, T. Mavrogordatos, T. S. Monteiro, and P. F. Barker, \emph{Cavity Cooling a Single Charged Levitated Nanosphere}, \href{https://doi.org/10.1103/PhysRevLett.114.123602}{Phys. Rev. Lett. \textbf{114}, 123602 (2015)}.
	
	\bibitem{dt1} Y. Leng, R. Li, X. Kong, H. Xie, D. Zheng, P. Yin, F. Xiong, T. Wu, C.-K. Duan, Y. Du, Z.-q. Yin, P. Huang, and J. Du, \emph{Mechanical Dissipation Below $1$ $\mu$Hz with a Cryogenic Diamagnetic Levitated Micro-Oscillator}, \href{https://doi.org/10.1103/PhysRevApplied.15.024061}{Phys. Rev. Applied \textbf{15}, 024061 (2021)}.
	
	\bibitem{dt2} D. Zheng, Y. Leng, X. Kong, R. Li, Z. Wang, X. Luo, J. Zhao, C.-K. Duan, P. Huang, J. Du, M. Carlesso, and A. Bassi, \emph{Room temperature test of the continuous spontaneous localization model using a levitated micro-oscillator}, \href{https://doi.org/10.1103/PhysRevResearch.2.013057}{Phys. Rev. Research \textbf{2}, 013057 (2020)}.
	
	\bibitem{microhertz} Y. Leng, R. Li, X. Kong, H. Xie, D. Zheng, P. Yin, F. Xiong, T. Wu, C.-K. Duan, Y. Du, Z. Yin, P. Huang, and J. Du, \emph{Mechanical Dissipation Below $1$ $\mu$Hz 
		with a Cryogenic Diamagnetic Levitated Micro-Oscillator}, \href{https://doi.org/10.1103/PhysRevApplied.15.024061}{Phys. Rev. Applied 15, 024061 (2021)}.

  	\bibitem{Chris} C. Whittle, E. D. Hall, S. Dwyer, N. Mavalvala, V. Sudhir, R. Abbott, A. Ananyeva, C. Austin, L. Barsotti, J. Betzwieser \textit{et al.}, \emph{Approaching the motional ground state of a 10-kg object}, \href{https://www.science.org/doi/10.1126/science.abh2634}{Science {\bf 372}, 1333 (2021)}.
	
	\bibitem{sql1} V. B. Braginsky, and F. Ya. Khalili, Quantum Measurements
	(Cambridge University Press, Cambridge, England, 1992).
	
	\bibitem{sql2} C. M. Caves, K. S. Thorne, R. W. P. Drever, V. D. Sandberg, and M. Zimmermann, \emph{On the measurement of a weak classical force coupled to a quantum-mechanical oscillator. I. Issues of principle}, \href{https://doi.org/10.1103/RevModPhys.52.341}{Rev. Mod. Phys. \textbf{52}, 341 (1980)}.
	
	\bibitem{sql3} A. Buikema \textit{et al.}, \emph{Sensitivity and performance of the Advanced LIGO detectors in the third observing run}, \href{https://doi.org/10.1103/PhysRevD.102.062003}{Phys. Rev. D 102, 062003 (2020)}.
	
	\bibitem{sql4} H. Yu, L. McCuller, M. Tse \textit{et al.}, \emph{Quantum correlations between light and the kilogram-mass mirrors of LIGO}, \href{https://doi.org/10.1038/s41586-020-2420-8}{Nature \textbf{583}, 43 (2020)}. 
	
	\bibitem{gonzalez2021levitodynamics} C. Gonzalez-Ballestero,  M. Aspelmeyer, L. Novotny, R.  Quidant, and O. Romero-Isart, \emph{Levitodynamics: Levitation and control of microscopic objects in vacuum}, \href{https://www.science.org/doi/10.1126/science.abg3027}{Science \textbf{374}, 6564 (2021)}.

 \bibitem{matsumoto1} N. Matsumoto, S. B. Catano-Lopez, M. Sugawara, S. Suzuki, N. Abe, K. Komori, Y. Michimura, Y. Aso, and K. Edamatsu, \emph{Demonstration of Displacement Sensing of a mg-Scale Pendulum for mm- and mg-Scale Gravity Measurements}, \href{https://doi.org/10.1103/PhysRevLett.122.071101}{Phys. Rev. Lett. \textbf{122}, 071101 (2019)}.

 \bibitem{matsumoto2} N. Matsumoto, and N. Yamamoto, \emph{Conditional mechanical squeezing of a macroscopic pendulum near quantum regimes}, \href{https://doi.org/10.48550/arXiv.2008.10848}{arXiv:2008.10848}
	

	
	\bibitem{sub-kelvin-Giesler} J. Gieseler, B. Deutsch, R. Quidant, and L. Novotny, \emph{Subkelvin Parametric Feedback Cooling of a Laser-Trapped Nanoparticle}, \href{https://doi.org/10.1103/PhysRevLett.109.103603}{Phys. Rev. Lett. {\bf 109}, 103603 (2012)}.
	
	\bibitem{Doherty-Jacobs} A. C. Doherty, and K. Jacobs, \emph{Feedback control of quantum systems using continuous state estimation}, \href{https://doi.org/10.1103/PhysRevA.60.2700}{Phys. Rev. A {\bf 60}, 2700 (1999)}.
	
	\bibitem{Daley} L. S. Walker, G. R. M. Robb, and A. J. Daley, \emph{Measurement and feedback for cooling heavy levitated particles in low-frequency traps}, \href{https://doi.org/10.1103/PhysRevA.100.063819}{Phys. Rev. A {\bf 100}, 063819 (2019)}.
	
	\bibitem{Ulbricht} A. Vinante, A. Pontin, M. Rashid, M. Toroš, P. F. Barker, and H. Ulbricht, \emph{Testing collapse models with levitated nanoparticles: Detection challenge}, \href{https://doi.org/10.1103/PhysRevA.100.012119}{Phys. Rev. A {\bf 100}, 012119  (2019)}.
	
	\bibitem{Magrini-Aspelmeyer} L. Magrini, P. Rosenzweig, C. Bach, A. Deutschmann-Olek, S. G. Hofer, S. Hong, N. Kiesel, A. Kugi, and M. Aspelmeyer, \emph{Real-time optimal quantum control of mechanical motion at room temperature},  \href{https://doi.org/10.1038/s41586-021-03602-3}{Nature {\bf 595}, 373 (2021)}.
	
	\bibitem{Tebbenjohanns-Novotny} F. Tebbenjohanns, M. Luisa Mattana, M. Rossi, M. Frimmer, and L. Novotny, \emph{Quantum control of a nanoparticle optically levitated in cryogenic free space}, \href{https://doi.org/10.1038/s41586-021-03617-w}{Nature {\bf 595}, 378 (2021)}.
	
	
	
	%\bibitem{Kurt} K. Jacobs, \emph{Quantum measurement theory and its applications}, Cambridge University Press (2014), 
	
	%\bibitem{Gerd} H. M. Wiseman,  and G. J. Milburn. \emph{Quantum measurement and control}. Cambridge University Press (2009).
	
	
	\bibitem{Vovrosh-Ulbricht} J. Vovrosh, M. Rashid, D. Hempston, J. Bateman, M. Paternostro, and H. Ulbricht, \emph{Parametric feedback cooling of levitated optomechanics in a parabolic mirror trap}, \href{https://doi.org/10.1364/JOSAB.34.001421}{J. Opt. Soc. Am. B {\bf 34}, 1421 (2017)}.
	
	\bibitem{Vahid} R. W. Taylor,  and V. Sandoghdar, \emph{Interferometric scattering microscopy: seeing single nanoparticles and molecules via Rayleigh scattering}, \href{https://doi.org/10.1021/acs.nanolett.9b01822}{Nano letters {\bf 19}, 4827 (2019)}.
	
	
	
	
	\bibitem{Tracy} G. Cerchiari, L. Dania, D. S. Bykov, R. Blatt, and T. E. Northup, \emph{Position measurement of a dipolar scatterer via self-homodyne detection}, \href{https://doi.org/10.1103/PhysRevA.104.053523}{Phys. Rev. A {\bf 104}, 053523 (2021)}.	
	
	\bibitem{HalliwellLatest} C. Mawby, and J. J.  Halliwell, \emph{Leggett-Garg violations for continuous variable systems with gaussian states}, \href{https://doi.org/10.48550/arXiv.2211.10292}{arXiv:2211.10292 [quant-ph]}.
	
\end{thebibliography}
\end{document}